\documentclass[journal]{IEEEtran}[10pt]
\IEEEoverridecommandlockouts
\usepackage{cite}
\usepackage{makecell}
\usepackage{amsmath,amssymb,amsfonts}
\usepackage{algorithmic}
\usepackage{bm}
\usepackage{graphicx}
\usepackage{subfigure}
\usepackage{makecell}
\newtheorem{theorem}{Theorem}

\newtheorem{corollary}{Corollary}[theorem]
\usepackage{textcomp}
\usepackage{epstopdf}
\usepackage{multirow}
\usepackage{xcolor}
\newtheorem{prop}{Proposition}
\def\BibTeX{{\rm B\kern-.05em{\sc i\kern-.025em b}\kern-.08em
		T\kern-.1667em\lower.7ex\hbox{E}\kern-.125emX}}
\begin{document}
	%\section{Rayleigh distance}
	\title{Cramér-Rao Bounds for Near-Field Sensing\\ with Extremely Large-Scale MIMO}
	\author{
	\IEEEauthorblockN{Huizhi~Wang, Zhiqiang~Xiao,~\IEEEmembership{Graduate Student Member, IEEE}, and Yong~Zeng,~\IEEEmembership{Senior Member, IEEE}}
	\thanks{This work was supported by the National Key R$\&$D Program of China with Grant number 2019YFB1803400. 
		 
	Part of this work has been presented at the 2022 IEEE ICC Workshops, Seoul, Korea
	in July 2022 \cite{b34}.
	
	The authors are with the National Mobile Communications Research Laboratory, Southeast University, Nanjing 210096, China. Y. Zeng and Z. Xiao are also with the Purple Mountain Laboratories, Nanjing 211111,China (e-mail: \{wanghuizhi, zhiqiang\_xiao, yong\_zeng\}@seu.edu.cn). (\emph{Corresponding author: Yong Zeng.})
	}}
	\maketitle
	
\begin{abstract}
\textbf{Mobile communication networks were designed to mainly support ubiquitous wireless communications, yet they are also expected to achieve radio sensing capabilities in the near future. However, most prior studies on radio sensing usually rely on far-field assumption with uniform plane wave (UPW) models. With the ever-increasing antenna size, together with the growing demands to sense nearby targets, the conventional far-field UPW assumption may become invalid. Therefore, this paper studies near-field radio sensing with extremely large-scale (XL) antenna arrays, where the more general uniform spheric wave (USW) sensing model is considered. Closed-form expressions of the Cramér-Rao Bounds (CRBs) for both angle and range estimations are derived for near-field XL-MIMO radar mode and XL-phased array radar mode, respectively. Our results reveal that different from the conventional UPW model where the CRB for angle decreases unboundedly as the number of antennas increases, for XL-MIMO radar-based near-field sensing, the CRB decreases with diminishing return and approaches to a certain limit as the number of antennas increases. Besides, different from the far-field model where the CRB for range is infinity since it has no range estimation capability, that for the near-field case is finite. Furthermore, it is revealed that the commonly used spherical wave model based on second-order Taylor approximation is insufficient for near-field CRB analysis. Extensive simulation results are provided to validate our derived CRBs. }
%	(may summarize the main insights from simulation results)
%	
%	We validate the derived CRBs with state-of-the-art localization algorithms and analyse the dependence of the CRBs on various system parameters.
\end{abstract}
	
\section{introduction}
With the fifth-generation (5G) mobile communication networks being commercially deployed, researchers have started the investigation of the key technologies for the sixth-generation (6G) networks~\cite{b26,b11,b10}. There is no doubt that 6G will continue to significantly improve the performance of wireless communications, in terms of coverage, connectivity density, data rate, latency, etc. On the other hand, it is also widely believed that 6G should go beyond communications, by providing various new services such as high-performance ubiquitous localization and radar sensing~\cite{b16,b17,b12}, which is possible thanks to the continuous expansion of cellular bandwidth and the ever-increasing of antenna size. Therefore, the integration of sensing and communication has received significant research interest recently, under various terms like joint communication and radar/radio sensing (JCAS)~\cite{b13}, dual-functional radar communications (DFRC)~\cite{b72}, and integrated sensing and communication (ISAC)~\cite{b20}.

Most of the research on ISAC can be loosely categorized into waveform design\cite{b59,b73,b60,b61,b71}, codebook design\cite{b62}, beam alignment ~\cite{b74}\cite{b76} and information-theoretical limits analysis\cite{b64,b65,b66}, etc. For radar sensing, several estimation-theoretic metrics such as Cramér-Rao Bound (CRB)\cite{b49}, Weiss-Wdinstein Bound~\cite{b50} and Ziv-Zakai Bound~\cite{b48} are used to evaluate the performance of parameter estimations, such as propagation delay, angle of arrival/departure (AoA/AoD), Doppler frequency, etc. Perhaps the most commonly used bound for parameter estimation is CRB, which serves as a lower bound for unbiased mean-square error (MSE) estimator. Different CRBs have been derived for two typical radar sensing modes, namely {\it MIMO radar mode} and {\it phased array radar mode}\cite{b6}. For MIMO radar mode, orthogonal waveforms are transmitted from different antennas, so as to obtain the waveform diversity gain. In this case, both colocated and distributed MIMO radar systems have been studied in terms of CRB analysis\cite{b28,b68,b69}. On the other hand, for phased array radar mode, coherent waveforms are transmitted from different antennas, so as to obtain high transmit coherent processing gain. The CRBs for monostatic phased array radar system with single transmit antenna and muti-antenna arrays have been studied in \cite{b51} and \cite{b67}, respectively. Existing results in~\cite{b28} reveal that for both MIMO and phased array radar modes, the CRBs for angle estimation decrease indefinitely with the increase of signal-to-noise ratios (SNRs) and the number of transmit and receive antennas. 

%Multiple-input multiple-output (MIMO) is a key enabling technology for both high-performance communication and radar sensing. It is widely known that MIMO communication offers the fundamental spatial multiplexing gain and diversity gain~\cite{b19}. There are similar gains for radar sensing with multiple antennas, corresponding to two different radar modes, namely {\it MIMO radar} and {\it phased array radar}\cite{b6}. For MIMO radar, orthogonal waveforms are transmitted from different antennas, so as to obtain the waveform diversity gain. On the other hand, for phased array radar, coherent waveforms are transmitted from multiple antennas, so as to obtain high transmit coherent processing gain. Compared with phased array radar, MIMO radar can obtain a larger virtual aperture by applying sparse array, which improves spatial resolution in angle measurement. On the other hand, phased array radar achieves higher beamforming gain by transmitting and receiving both with narrow beams, but its sensing area during each pulse is limited.

On the other hand, MIMO communications have been tremendously advanced from small MIMO in 4G to massive MIMO in 5G\cite{b21}. Looking forward towards 6G, there have been growing interests in the study of extremely large-scale MIMO (XL-MIMO)\cite{b2,b29,b23,b3,b24}, for which the antenna size is so large that conventional far-field assumption with uniform plane wave (UPW) models become invalid. Instead, the more generic spherical wavefront characteristics need to be taken into account\cite{b34}. However, most existing studies mentioned above for CRB analysis mainly rely on the conventional UPW models\cite{b25}, which was justifiable since most prior radar sensing applications were mainly for distant targets and the antenna size is usually moderate. With the ever-increasing antenna size at base stations (BSs), together with the growing demands to also sense nearby targets, it is necessary to develop new CRB analysis for near-field sensing, without restricting to the conventional far-field UPW models.

There are some relevant works for CRB analysis that consider the near field effect for source localization problems\cite{b52,b53,b54,b55}. For example, Fresnel approximation based on second-order Taylor approximation is commonly used to approximate spherical wavefront\cite{b52}\cite{b53}. In \cite{b1}, a near-field tracking problem for inferring the position and velocity of a moving source was considered, and the posterior Cramér-Rao Lower Bound was derived. However, although such a second-order Taylor approximation method well fits the exact near-field uniform spheric wave (USW) model in most practical systems, it may introduce some systematic errors and make the model asymptotically biased\cite{b56}. Moreover, most existing results are derived for the source localization problem that involves only one-hop signal propagation, which cannot be applied for radar sensing scenario with double-hop signal propagation. In~\cite{b54}, the authors derived the conditional and unconditional CRBs for near-field bistatic MIMO radar system. However, such results were dependent on the derivative of the path difference with respect to unknown parameters, which is difficult to gain insights between the CRBs and the key system parameters or array configuration. To the best of our knowledge, closed-form CRB expressions in terms of the key system parameters, such as SNR and number of antennas, have not been reported for near-field radar sensing taking into account uniform spherical wave (USW) characteristics. This motivates our current work. The main contributions of this paper are summarized as follows:
\begin{itemize}
	\item First, we present the near-field bistatic sensing model with extremely large-scale antenna arrays, for which the signal processing procedures for XL-MIMO radar mode and XL-phased array radar mode are introduced, respectively. It is found that directly deriving the CRBs for near-field bistatic sensing is challenging, since it involves four-dimension parameter estimation, including transmitter-side and receiver-side angles and ranges, respectively. To tackle this difficulty, we transform the problem into two-dimensional parameter estimation problem by exploiting the geometrical relationship between the transmit and receive arrays, so that the receiver-side parameters can be represented in terms of the transmitter-side parameters. 
	\item Next, to gain useful insights, the basic monostatic near-field sensing is first considered, which can be viewed as a special case of the general bistatic near-field sensing. The closed-form expressions of the near-field CRBs for angle and range estimation are derived. The asymptotic cases with very large target range or antenna size are respectively considered to gain useful insights. It is revealed that our newly derived CRBs for near-field USW-based sensing include the results based on the conventional far-field UPW model as special cases. Different from the conventional far-field sensing, for XL-MIMO near-field sensing, the CRB for angle estimation no longer decreases indefinitely as the array size increases. Instead, it would approach to a limit that is dependent on the inter-element spacing. Moreover, the CRB for range estimation, which is infinity in conventional UPW model, is shown to be finite in the near-field case, showing the capability for range discrimination with XL-MIMO near-field sensing.
	\item Finally, for the more general bistatic XL-MIMO sensing, as it is quite challenging to derive the closed-form CRB expressions when near-field USW model is considered at both transmitter and receiver sides, we consider the more tractable and likely scenario in practice that the sensing target locates in the near-field of the transmit array. The corresponding closed-form CRBs of angle and range are derived and useful insights are obtained. Furthermore, by comparing our CRBs with the classic Capon algorithm~\cite{b81}, numerical results are provided to validate our derived near-field CRBs.
\end{itemize}

The rest of this paper is organized as follows. Section \uppercase\expandafter{\romannumeral2} introduces the general near-field USW model for bistatic radar sensing, together with the key radar signal processing procedures for XL-MIMO radar mode and XL-phased array radar mode, respectively. Section \uppercase\expandafter{\romannumeral3} derives the closed-form expression of CRB for the monostatic scenario, which can be treated as a special case of bistatic sensing. Section \uppercase\expandafter{\romannumeral4} derives the CRBs for bistatic sensing when the sensing target is located at the near field of the transmit array. Section \uppercase\expandafter{\romannumeral5} provides numerical results to validate our derived CRBs.  

\textit{Notations}: Lower and upper-case bold letters denote vectors and matrices, respectively. ${\bf{z}}_i$ denotes the $i$-th element of a vector ${\bf z}$. ${\bf{Z}}^T$, ${\bf{Z}}^*$, ${\bf{Z}}^H$ and$\ \det ({\bf{Z}})\ $denote the transpose, conjugate, conjugate transpose, and determinant of the matrix ${\bf{Z}}$, respectively. $\mathcal{R\{\cdot\}}$ denotes the real part, and$\ \otimes \ $denotes the Kronecker
product.$\ \frac{\partial }{{\partial z}}\left(  \cdot  \right)\ $denotes the partial derivative of $z$.$\ {{\bf{1}}_L}\ $denotes the vector of dimension$\ L \times 1\ $with all ones. Finally,
$j$ denotes the imaginary unit and$\ {\| {\bf{z}} \|}$ denotes the Euclidean norm of vector $\ {\bf{z}}\ $.

\begin{figure}[htbp]
%	\centering
%	\vspace{-3.5ex}
	\setlength{\abovecaptionskip}{-0.1cm}
	\setlength{\belowcaptionskip}{-0.3cm}
	\centerline{\includegraphics[width=0.45\textwidth]{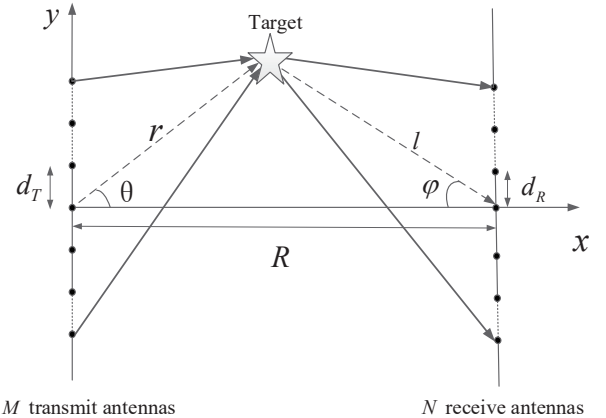}}
	\caption{Near-field radar sensing with XL-MIMO. }
	\label{fig2}
\end{figure}
\section{System model}
As shown in Fig.\ref{fig2}, we consider a near-field radar sensing system with XL-MIMO. Let $M\gg 1$ and $N\gg 1$ denote the number of transmit and receive antenna elements, respectively. For notational convenience, we assume that $M$ and $N$ are odd numbers. Furthermore, both the transmitter and receiver are equipped with uniform linear arrays (ULAs) with inter-element spacing denoted by $d_T$ and $d_R$, respectively. Thus, the array apertures of the transmitter and receiver are$\ {D_T} \approx M{d_T}\ $and$\ {D_R} \approx N{d_R},\ $respectively. For simplicity, we assume that the transmit and receive ULAs are parallel to each other and their distance is $R$. Without loss of generality, the transmit ULA is placed along the $y$-axis and centered at the origin. Therefore, the location of the $m$th transmit element is$\ {{\bf{w}}_m} = {[0,m{d_T}]^T},\ $where$\ m \in \mathcal{M}$, with $\mathcal{ M }  \triangleq  \left\{ {0, \pm 1, \cdots , \pm {{(M - 1)} \mathord{\left/
			{\vphantom {{(M - 1)} 2}} \right.
			\kern-\nulldelimiterspace} 2}} \right\}.\ $Similarly, the location of the $n$th receive element is $[R,n{d_R}]^T,\ $where$\ n \in \mathcal{N}$, with $\mathcal{ N }  \triangleq  \left\{ {0, \pm 1, \cdots , \pm {{(N - 1)} \mathord{\left/
			{\vphantom {{(N - 1)} 2}} \right.
			\kern-\nulldelimiterspace} 2}} \right\}.\ $Let$\ {\bf{q}} = {[r\cos \theta ,r\sin \theta ]^T}\ $denotes the location of the radar sensing target, where $r$ is the distance between the target and the center of the transmit array, and$\ \theta  \in \left[ -\frac{\pi}{2},\frac{\pi}{2} \right]\ $is the direction of the target with respect to the normal vector of the transmit array. Therefore, the distance between the target and the $m$th transmit antenna is 
\begin{equation}
	\setlength\abovedisplayskip{2pt}
	\setlength\belowdisplayskip{2pt}
\footnotesize
	\begin{aligned}
		{r_m} = \| {{{\bf{w}}_m}{\rm{ - }}{\bf{q}}} \| 
		&= r\sqrt {1 - 2m{\varepsilon _T}\sin \theta  + {m^2}{\varepsilon _T}^2} ,
	\end{aligned}
	\label{eq9}
\end{equation}
where$\ {\varepsilon _T} \triangleq\frac{{{d_T}}}{r} \ll 1.\ $Note that (\ref{eq9}) is the exact distance expression that can be degenerated to the conventional far-field UPW model by using first-order Taylor approximation when $D_T\ll r$.

For XL-MIMO systems, when the far-field assumption no longer holds, the exact distance expression (\ref{eq9}) is usually needed to accurately model the signal phase and amplitude variations across different array elements. 
In this case, the element of the transmit array response vector not only depends on the direction $\theta$, but also on the range $r$, which can be expressed as$\ {\tilde a_m}(r,\theta ) = \frac{{\sqrt {{\alpha _0}} }}{{{r_m}}}{e^{ - j\frac{{2\pi }}{\lambda }{r_m}}}$\cite{b2}, $m\in \mathcal M$, with$\ {{\alpha_0}}\ $denoting the channel power gain at the reference distance of $1$m.

Similarly, let $l$ denotes the distance between the target and the center of the receive antenna array, and $\varphi$ denotes the direction of the target with respect to the normal vector of the receive array. Therefore, the element of the receive array response vector can be expressed as
$\ {\tilde b_n}(l,\varphi ) = \frac{{\sqrt {{\beta _0}} }}{{{l_n}}}{e^{ - j\frac{{2\pi }}{\lambda }{l_n}}}, n \in \mathcal{N},$ with$\ {l_n} = l\sqrt {1 - 2n{\varepsilon _R}\sin  \varphi + {n^2}{\varepsilon _R}^2}\ $denoting the distance between the target and the center of the receive array.$\ {\varepsilon _R} \buildrel \Delta \over = \frac{{{d_R}}}{l} \ll 1,\ $and$\ {{\beta_0}}\ $denotes the channel power gain at the reference distance of $1$m. Furthermore, when the distance $R$ between the transmit and receive arrays is known, the receiver side range and angle parameters $l$ and $\varphi$ can be expressed in terms of the transmitter side parameters $r$ and $\theta$, i.e.,
\begin{equation}
\setlength\abovedisplayskip{1.5pt}
	\setlength\belowdisplayskip{1.5pt}
	\footnotesize
	\hspace{-5ex}
\ \begin{aligned}
			&l( {r,\theta } ) =  \sqrt {{R^2} + {r^2} - 2Rr\cos \theta } ,\\
			&\varphi (r,\theta ) = \arcsin \left\{ {\frac{{r\sin \theta }}{{\sqrt {{R^2} + {r^2} - 2Rr\cos \theta } }}} \right\}.\
\end{aligned}\
	\label{transform}
\end{equation}
As a result, the distance $l_n$ and the element of the receive array response vector ${\tilde b_n}(l,\varphi)$ can be represented in terms of $r$ and $\theta$ as
\begin{equation}
	\setlength\abovedisplayskip{2pt}
	\setlength\belowdisplayskip{2pt}
	\footnotesize
%	\hspace{-1ex}
	\ \begin{aligned}
	\ {l_n}( {r,\theta } ) &= \sqrt {{R^2} + {r^2} - 2Rr\cos \theta  - 2n{d_R}r\sin \theta  + {n^2}{d_R}^2} ,\\
	\ {\tilde b_n}(r,\theta ) &= \frac{{\sqrt {{\beta _0}} }}{{{l_n}(r,\theta )}}{e^{ - j\frac{{2\pi }}{\lambda }{l_n}(r,\theta )}}.
	\end{aligned}\
\label{t}
\end{equation}

Let $x_m(t)$ denotes the transmitted waveform by the $m$th transmit antenna, $m\in \mathcal M$. The received signal by the $n$th receive antenna due to target reflection can be expressed as
\begin{equation}
			\setlength\abovedisplayskip{2pt}
	\setlength\belowdisplayskip{2pt}
	\footnotesize
	\begin{aligned}
	\ {r_n}(t) = \tilde \kappa {\tilde b_n}(r,\theta )\sum\nolimits_{m =  - \frac{{M - 1}}{2}}^{\frac{{M - 1}}{2}} {{\tilde a_m}(r,\theta ){x_m}(t - {\tau}) + {n_n}(t)}, \
	\end{aligned}
	\label{re1}
\end{equation}
where $\tilde{\kappa}$ is a complex reflection coefficient that includes the impact of radar cross section (RCS) of the target, $\tau$ is the propagation delay of the reflected signal by the target. Note that we assume that the propagation delays between different transmit and receive elements are approximately equal, which is valid when $D_T+D_R\le \frac{c}{B},\ $where $B$ denotes system bandwidth, and $c$ is the speed of light. $n_n(t)$ is the independent and identically distributed (i.i.d.) additive white Gaussian noise (AWGN) with power spectral density $N_0$.

Note that$\ {\tilde a_m}(r,\theta )\ $can be equivalently written as$\ {\tilde a_m}(r,\theta ) = \frac{{\sqrt {{\alpha _0}} }}{r}\frac{r}{{{r_m}}}{e^{ - j\frac{{2\pi }}{\lambda }{r_m}}},\ $and when $r>1.2D_T$\cite{b40}, the amplitude variations across array elements can be neglected. Therefore, the transmit array response vector can be expressed as ${\tilde{ \bf a}}(r,\theta ) = \frac{{\sqrt {{\alpha _0}} }}{r}{\bf{a}}(r,\theta ),$ where
\begin{equation}
	\setlength\abovedisplayskip{2pt}
	\setlength\belowdisplayskip{2pt}
%		\hspace{-1.3ex}
\footnotesize
	\begin{aligned}
{\bf{a}}(r,\theta) = [{a_{ - \frac{{M - 1}}{2}}}(r,\theta),...,{a_m}(r,\theta),...,{a_{ \frac{{M - 1}}{2}}}(r,\theta)]^T,
	\end{aligned}
	\label{eq2}
\end{equation}
with the element$\ {a_m}(r,\theta ) = {e^{ - j\frac{{2\pi }}{\lambda }{r_m}}}.\ $Similarly, the receive response vector can be expressed as ${\tilde {\bf b}}(r,\theta )
= \frac{{\sqrt {{\beta _0}} }}{l}{\bf{b}}(r,\theta ),$ where
\begin{equation}
	\setlength\abovedisplayskip{2pt}
	\setlength\belowdisplayskip{2pt}
%		\hspace{-1.ex}
\footnotesize
	\begin{aligned}
	{\bf{b}}(r,\theta) = [{b_{ - \frac{{N - 1}}{2}}}(r,\theta),...,{b_n}(r,\theta),...,{b_{ \frac{{N - 1}}{2}}}(r,\theta)]^T,
	\end{aligned}
	\label{eq3}
\end{equation}
with$\ {b_n}(r,\theta ) = {e^{ - j\frac{{2\pi }}{\lambda }{l_n}}}.\ $Note that we have expressed the receive array response vector in terms of the transmitter side angle and range parameters $(r,\theta)$ based on the relationship (\ref{t}). 
Therefore, according to (\ref{re1}), the vector form of the received signal for bistatic near-field radar sensing can be written as
	\begin{equation}
		\setlength\abovedisplayskip{2pt}
		\setlength\belowdisplayskip{1.5pt}
		\footnotesize
		\begin{aligned}
		\ {\bf{r}}(t) = \kappa {\bf{b}}(r,\theta ){{\bf{a}}^T}(r,\theta ){\bf{x}}(t - \tau ) + {\bf{n}}(t),\	
		\end{aligned}
		\label{eq11}
	\end{equation}
where$\ {\bf{x}}(t)=[x_m(t)]_{m\in \mathcal M} \ $denotes the transmitted waveform vector,$\ \kappa  \triangleq \tilde \kappa \frac{\sqrt{\alpha_0\beta_0} }{{r\sqrt {{R^2} + {r^2} - 2Rr\cos \theta } }}\ $is the coefficient taking into account the reference power gains,$\ {\bf{n}}(t)\in \mathbb{C}^{N\times 1}$ is the i.i.d. AWGN with zero mean and power spectral density$\ N_0.\ $Note that the coefficient $\kappa$ also depends on the target location $(r,\theta)$ in general. However, since the variation of amplitude is much less sensitive than the phase variation, we ignore the dependence of $\kappa$ on $(r,\theta)$ for the subsequent CRB derivation.

%Besides, for monostatic scenario, the expression of received signal can be obtained by simply substituting $R=0$ and$\ {\bf{b}}(r,\theta ) = {\bf{a}}(r,\theta )\ $into (\ref{eq11}). Therefore, we use (\ref{eq11}) as the universal expression for both monostatic and bistatic scenario in the following part of this paper.

In the following, we consider two standard radar modes, i.e., MIMO radar mode and phased array radar mode\cite{b6}, which we term as {\it XL-MIMO radar} and {\it XL-phased array radar} respectively in the context of near-field sensing with extremely large-scale antenna arrays\cite{b34}. 
\subsection{XL-MIMO Radar}
	For MIMO radar, the transmitted waveform $\mathbf x(t)$ in (\ref{eq11}) is
	\begin{equation}
		\setlength\abovedisplayskip{1pt}
		\setlength\belowdisplayskip{2pt}
		\footnotesize
		\ {{\bf{x}}}(t) = \sqrt {\frac{P}{M}} {\bf{s}}(t),\
		\label{eq18}
	\end{equation}
where $P$ is the total transmit power, and$\ {\bf{s}}(t)=[s_m(t)]_{m\in \mathcal M} \ $represents the $M$ orthogonal waveforms, which satisfy\cite{b9}
	\begin{equation}
		\setlength\abovedisplayskip{2pt}
		\setlength\belowdisplayskip{1pt}
		\begin{aligned}
\ \frac{1}{{{T_p}}}\int_{{T_p}} {{s_m}(t)s_k^*(t - \alpha )} dt = \begin{cases}
	{{R_{ss}}(\alpha )},& \text{ $ m=k, $ } \\
	0,& \text{ $ m \ne k,$ }
\end{cases}
		\end{aligned}
	\end{equation}
where $T_p$ is the duration of coherent processing interval (CPI), and$\ {R_{ss}}(\alpha)\ $is the autocorrelation function of the waveforms $s_m(t)$, with ${R_{ss}(0)}=1$. By substituting (\ref{eq18}) into (\ref{eq11}), the received signal for XL-MIMO radar is\cite{b34}
	\begin{equation}
		\setlength\abovedisplayskip{1.5pt}
		\setlength\belowdisplayskip{1.5pt}
		%\hspace{-1.8ex}
\footnotesize
			\begin{aligned}
				{\bf{r}}(t) &= \kappa \sqrt {\frac{P}{M}} {\bf{b}}(r,\theta ){{\bf{a}}^T}(r,\theta ){\bf{s}}(t - \tau) + {\bf{n}}(t)\\
				&= \kappa \sqrt {\frac{P}{M}} {\bf{b}}(r,\theta )\sum\nolimits_{m =  - \frac{{M - 1}}{2}}^{\frac{{M - 1}}{2}} {{a_m}(r,\theta ){s_m}(t - \tau)}  + {\bf{n}}(t).
		\end{aligned}
		\hspace{-1.8ex}
	\end{equation}
	By applying matched filtering to ${\bf{r}}(t)$ with each of the orthogonal waveforms$\ s_k(t - \alpha),\ $$k\in \mathcal {M}$, where $\alpha$ is some selected time delay that may be different from the groundtruth delay $\tau$, the output signal can be expressed as 
	\begin{equation}
		\setlength\abovedisplayskip{2pt}
		\setlength\belowdisplayskip{1pt}
		\footnotesize
			\begin{aligned}
				{{\bf{y}}_{k}} &= \frac{1}{\sqrt{T_p}}\int_{{T_p}} {{\bf{r}}(t)} s_k^*(t - \alpha)dt\\
				&=\kappa \sqrt {\frac{{T_p}P}{M}} {\bf{b}}(r,\theta ){a_k}(r,\theta ){R_{ss}}(\alpha  - \tau ) + {\tilde {\bf{n}}_k},
		\end{aligned}
	\end{equation}
where$\ {{{\bf{\tilde n}}}_k} \buildrel \Delta \over = \frac{1}{{\sqrt{T_p}}}\int_{{T_p}} {{\bf{n}}(t)} s_k^*(t - \alpha)dt.\ $The normalization factor$\ \frac{1}{{\sqrt {{T_p}} }}\ $is applied to ensure that the noise remains to have variance $N_0$. By concatenating $\mathbf y_{k}\in \mathbb{C}^{N\times 1}$ for all $k\in \mathcal {M}$, and if the matched filter delay $\alpha$ mathces with the groundtruty delay $\tau$, we obtain the following $MN$ dimensional data vector
	\begin{equation}
		\setlength\abovedisplayskip{0.5pt}
		\setlength\belowdisplayskip{1.5pt}
		\footnotesize
		\begin{aligned}
			\ {\bf{y}} = \kappa \sqrt {\frac{{T_p}P}{M}} {\bf{b}}(r,\theta ) \otimes {\bf{a}}(r,\theta ) + {\bf{\tilde n}},\
		\end{aligned}
		\label{eq5}
	\end{equation}
where$\ {\bf{\tilde n}}=[\tilde{\mathbf n}_k]_{k\in \mathcal M}\in \mathbb{C}^{MN\times 1} $represents the resulting noise after matched-filtering, which can be shown to have zero mean and variance $N_0$.
\subsection{XL-phased Array Radar}
For phased array radar, transmit beam is formed to search or track the target at a certain direction $\theta'$ and range $r'$\cite{b34}. In this case, the transmitted signal in (\ref{eq11}) is
	\begin{equation}
		\setlength\abovedisplayskip{1pt}
		\setlength\belowdisplayskip{1.5pt}
		\footnotesize
		\ {\bf{x}}(t) = \frac{{\sqrt P }}{{\| {{\bf{a}}(r,\theta )} \|}}{{\bf{a}}^*}(r',\theta ')s(t),\
		\label{eq12}
	\end{equation}
where $P$ is the total transmit power,$\ {\bf{a}}(r',{\theta'})\ $is the transmit steering vector towards the target at range $r'$ and angle $\theta',$ $s(t)$ is the single transmitted waveform satisfying$\ \frac{1}{{{T_p}}}\int_{{T_p}} {s(t)} {s^*}(t - \alpha )dt = R(\alpha ),\ $where $R(\alpha)$ is the autocorrelation function for phased array radar.
By substituting (\ref{eq12}) into (\ref{eq11}), the received signal for XL-phased array radar is
	\begin{equation}
		\setlength\abovedisplayskip{1pt}
		\setlength\belowdisplayskip{2pt}
		%\hspace{1.4ex}
		\footnotesize
		\begin{aligned}
		{\bf{r}}(t) = \kappa {\frac{{\sqrt P }}{{\| {{\bf{a}}(r,\theta )} \|}}}{\bf{b}}(r,\theta ){{\bf{a}}^T}(r,\theta ){{\bf{a}}^*}(r',\theta ')s(t - \tau ) + {\bf{n}}(t).\
		\end{aligned}
	\end{equation}
By applying matched filtering for$\ {\bf{r}}(t)\ $with the transmitted waveform$\ s(t - \alpha ),\ $we have
\begin{equation}
		\setlength\abovedisplayskip{2pt}
		\setlength\belowdisplayskip{2pt}
		%\hspace{-1ex}
		\footnotesize
\begin{aligned}
	{\bf{y}}(\alpha ,r',\theta ') &= \frac{1}{{\sqrt{T_p}}}\int_{{T_p}} {{\bf{r}}(t){s^*}(t - \alpha )} dt\\
	 &= \kappa \frac{{\sqrt {{T_p}P} }}{{\| {{\bf{a}}(r,\theta )} \|}}{\bf{b}}(r,\theta ){{\bf{a}}^T}(r,\theta ){{\bf{a}}^*}(r',\theta ')R(\alpha  - \tau ) + {\bf{\tilde n}},
\end{aligned}
\label{ph}
	\end{equation}
where$\ {\bf{\tilde n}} \buildrel \Delta \over = \frac{1}{{\sqrt{T_p}}}\int\limits_{{T_P}} {{\bf{n}}(t-\alpha)} {s^*}(t)dt\ $is the resulting noise vector with zero mean and variance $N_0$. When the searching parameters match with the groundtruth values, i.e., $\theta' = \theta,$ $r' = r,$ $\alpha = \tau ,\ $and by noting that $\|{\bf a}(r,\theta)\|=\sqrt{M}$, we have
	\begin{equation}
	\setlength\abovedisplayskip{2pt}
	\setlength\belowdisplayskip{0pt}
	%\hspace{-1ex}
	\footnotesize
		\begin{aligned}
		 {\bf{y}} = \kappa \sqrt {{T_p}MP} {\bf{b}}(r,\theta ) + {\bf{\tilde n}}.
		\label{xlph}
		\end{aligned}
	\end{equation}
\subsection{Cramér-Rao Bound}\label{crb}
It follows from (\ref{eq5}) and (\ref{xlph}) that, for both XL-MIMO radar mode and XL-phased array radar mode, the resulting signal after matched filter can be written in the unified form as
	\begin{equation}
		\setlength\abovedisplayskip{2pt}
		\setlength\belowdisplayskip{2pt}
		\footnotesize
		\ {\bf{y}} = \rho {\bf{g}} + {\bf{\tilde n}},\
		\label{model}
	\end{equation}
where $\rho$ is a constant that is approximately independent of the sensing parameters $\theta$ and $r$. For XL-MIMO radar mode, we have$\ {\bf{g}} = {\bf{b}}(r,\theta ) \otimes {\bf{a}}(r,\theta )\ $and $ \rho  = \kappa \sqrt {\frac{{T_p}P}{M}},\ $while for XL-phased array radar mode, we have$\ {\bf{g}} = {\bf{b}}(r,\theta )\ $and$\ \rho  = \kappa \sqrt {{T_p}PM}\ $. 
	
Let$\ {\bf{w}} = \rho {\bf{g}}\ $and$\ {\bf{z}} = {[\theta ,{r},{\kappa _r},{\kappa _i}]^T}\ $that includes the unknown parameters, where$\ \kappa_r\ $and$\ \kappa_i \ $denote the real and imaginary parts of $\kappa$, respectively.
According to\cite{b28}, the Fisher's information matrix (FIM) with respect to $\bf{z}$ can be expressed as
\begin{equation}
	\setlength\abovedisplayskip{0.5pt}
	\setlength\belowdisplayskip{1pt}
	\footnotesize
	\begin{aligned}
	{\bf{F}} & = \frac{2}{{{N_0}}}\Re \left\{ {\left( {\frac{{\partial {\bf{w}}}}{{\partial {\bf{z}}}}} \right){{\left( {\frac{{\partial {\bf{w}}}}{{\partial {\bf{z}}}}} \right)}^H}} \right\}\\ 
&= \frac{2}{{N_0}}\left[ {\begin{array}{*{20}{c}}
		{{v_{\theta \theta }}}&{{v_{\theta r}}}&\vline& {{v_{\theta\kappa_r }}}&{{v_{\theta \kappa_i }}}\\
		{{v_{\theta r}}}&{{v_{rr}}}&\vline& {{v_{r\tilde \kappa }}}&{{v_{r\kappa_i }}}\\
		\hline
		{{v_{\theta\kappa_r }}}&{{v_{r\kappa_r }}}&\vline& {{v_{\kappa_r\kappa_r }}}&0\\
		{{v_{\theta \kappa_i }}}&{{v_{r\kappa_i }}}&\vline& 0&{{v_{\kappa_i \kappa_i }}}
\end{array}} \right] = \left[ {\begin{array}{*{20}{c}}
		{{{\bf{\Pi }}_{11}}}&\vline& {{{\bf{\Pi }}_{12}}}\\
		\hline
		{{{\bf{\Pi }}_{21}}}&\vline& {{{\bf{\Pi }}_{22}}}
\end{array}}\right ],
	\end{aligned}
\end{equation}
where $v_{z_1z_2}\triangleq\Re \{ {( {\frac{{\partial {\bf{w}}}}{{\partial {z_1}}}} ){( {\frac{{\partial {\bf{w}}}}{{\partial {z_2}}}} )^H}}\}. $ 
The CRB for the parameters of interest $(r,\theta)$ is related to the inverse of the FIM
	\begin{equation}
	\setlength\abovedisplayskip{0.5pt}
	\setlength\belowdisplayskip{1.5pt}
	\footnotesize
	\begin{aligned}
		\ {{\bf{F}}^{ - 1}} = \frac{{{N_0}}}{2}\left[ {\begin{array}{*{20}{c}}
			{{{\bf{Q}}^{ - 1}}}&\vline&  \times \\
			\hline
			\times &\vline&  \times 
	\end{array}} \right],\
	\end{aligned}
\end{equation}
where$\ {\bf{Q}}{\rm{ = }}{{\bf{\Pi }}_{11}} - {{\bf{\Pi }}_{12}}{\bf{\Pi }}_{22}^{ - 1}{\bf{\Pi }}_{12}^T\ $is the Schur complement of$\ {{\bf{\Pi }}_{22}}\ $corresponding to$\ {\bf{F}}.\ $ Furthermore, it is shown in \cite{b28} that$\ {\bf{Q}} = {| \rho  |^2}{\bf{Q}}',\ $with
	\begin{equation}
	\setlength\abovedisplayskip{0.5pt}
	\setlength\belowdisplayskip{1.5pt}
	\footnotesize
	\begin{aligned}
		\ {\bf{Q}}' = \left[ {\begin{array}{*{20}{c}}
			{{{\| {{{\bf{g}}_\theta }} \|}^2}{{\sin }^2}\Omega }&{\frac{{\mathfrak{R}\{ {{{\bf{g}}^H}{\bf{Wg}}} \}}}{{{{\| {\bf{g}} \|}^2}}}}\\
			{\frac{{\mathfrak{R}\{ {{{\bf{g}}^H}{\bf{Wg}}} \}}}{{{{\| {\bf{g}} \|}^2}}}}&{{{\| {{{\bf{g}}_r}} \|}^2}{{\sin }^2}\Theta }
	\end{array}}\right ],\
	\end{aligned}
\end{equation}
where$\ {{\bf{g}}_\theta } = \frac{{\partial {\bf{g}}}}{{\partial \theta }},\ $$\ {{\bf{g}}_r} = \frac{{\partial {\bf{g}}}}{{\partial r}},\ $ $\ {\sin ^2}\Omega  = 1 -  {\frac{{{{| {{\bf{g}}_\theta ^H{\bf{g}}} |}^2}}}{{{{\| {{{\bf{g}}_\theta }} \|}^2}{{\| {\bf{g}} \|}^2}}}} ,\ $$\ {\sin ^2}\Theta  = 1 -  {\frac{{{{| {{\bf{g}}_r^H{\bf{g}}} |}^2}}}{{{{\| {{{\bf{g}}_r}} \|}^2}{{\| {\bf{g}} \|}^2}}}},\ $and$\ {\bf{W}} = ( {{\bf{g}}_\theta ^H{{\bf{g}}_r}} ){\bf{I}} - {{\bf{g}}_\theta }{\bf{g}}_r^H.\ $
	
Therefore, the CRBs of angle $\theta$ and range $r$ can be expressed as\cite{b28}
\begin{equation}
	\setlength\abovedisplayskip{0pt}
	\setlength\belowdisplayskip{0pt}
	%\hspace{-1ex}
\footnotesize
	\begin{aligned}
	\ CR{B_\theta } = \frac{{{N_0}}}{{2{{| \rho  |}^2}}}\frac{{{{\| {{{\bf{g}}_r}} \|}^2}{{\sin }^2}\Theta }}{{\det {{\bf{Q}}^\prime }}},\
		\label{theta}
	\end{aligned}
\end{equation}
\begin{equation}
	%\hspace{-1ex}
		\setlength\abovedisplayskip{0pt}
	\setlength\belowdisplayskip{-1pt}
\footnotesize
	\begin{aligned}
\ CR{B_r} = \frac{{{N_0}}}{{2{{| \rho  |}^2}}}\frac{{{{\| {{{\bf{g}}_\theta }} \|}^2}{{\sin }^2}\Omega }}{{\det {\bf{Q}}'}}.\ 
\label{dis}
	\end{aligned}
\end{equation}
Thus, the remaining task for the near-field CRB derivation for the input-output relation (\ref{model}) is to obtain the terms ${{\| {{{\bf{g}}_r}} \|}^2}$, ${{\| {{{\bf{g}}_\theta}} \|}^2}$, ${\det {\bf{Q}}'}$, ${{\sin }^2}\Theta$ and ${{\sin }^2}\Omega$ appearing in (\ref{theta}) and (\ref{dis}). In the following, to gain useful insights, we first consider near-field sensing for the basic monostatic scenario, which can be treated as a special case of the bistatic setup in Fig. \ref{fig2}, by letting $R=0$, $M=N$, $d_T=d_R$, and ${\bf{b}} (r,\theta)={\bf{a} }(r,\theta)$. After that, the more complicated bistatic sensing scenario will be considered in Section \ref{bistaticsensing}.
\section{Near-Field CRB for Monostatic Sensing}{\label{mono}}
%	\begin{figure}[htbp]
%%		\vspace{-3.5ex}
%		\setlength{\abovecaptionskip}{-0.1cm}
%		\setlength{\belowcaptionskip}{-0.3cm}
%		\centerline{\includegraphics[scale=0.5]{monostatic.eps}}
%		\caption{Monostatic radar sensing with extremely large-scale MIMO. }
%		\label{monos}
%%		\vspace{-2ex} 
%	\end{figure}
\subsection{XL-MIMO Radar}\label{monomimo}
In order to derive the closed-form expressions for the angle and range CRBs in (\ref{theta}) and (\ref{dis}), the terms$\ \| {{{\bf{g}}_\theta }} \|^2,\| {{{\bf{g}}_r}} \|^2,\| {\bf{g}} \|^2,\Re \{ {{{\bf{g}}^H}{\bf{Wg}}} \},{\sin ^2}\Omega ,\ $and$\ {\sin ^2}\Theta \ $should be derived. For the special case of monostatic sensing where ${\bf{b}} (r,\theta)={\bf{a} }(r,\theta)$, we have$\ {\bf{g}} = {\bf{a}}(r,\theta ) \otimes {\bf{a}}(r,\theta ).\ $Thus, we can derive the following results:
\begin{equation}
	\setlength\abovedisplayskip{1pt}
	\setlength\belowdisplayskip{0pt}
%	\hspace{-2.8ex}
\footnotesize
	\begin{aligned}
				{\| {{{\bf{g}}_\theta }} \|^2} &= \left[ {\frac{{\partial {{\bf{a}}^H}(r,\theta )}}{{\partial \theta }} \otimes {{\bf{a}}^H}(r,\theta ) + {{\bf{a}}^H}(r,\theta ) \otimes \frac{{\partial {{\bf{a}}^H}(r,\theta )}}{{\partial \theta }}} \right]\\
		&\times \left[ {\frac{{\partial {\bf{a}}(r,\theta )}}{{\partial \theta }} \otimes {\bf{a}}(r,\theta ) + {\bf{a}}(r,\theta ) \otimes \frac{{\partial {\bf{a}}(r,\theta )}}{{\partial \theta }}} \right]\\
		&= 2{\left\| {\frac{{\partial {\bf{a}}(r,\theta )}}{{\partial \theta }}} \right\|^2}{\left\| {{\bf{a}}(r,\theta )} \right\|^2} + 2{\left| {\frac{{\partial {{\bf{a}}^H}(r,\theta )}}{{\partial \theta }}{\bf{a}}(r,\theta )} \right|^2}\\
		&= 2Ma + 2{| c |^2},
	\end{aligned}
\label{parameter1}
\end{equation}
where$\footnotesize {\ a  \triangleq  {\left\| {\frac{{\partial {\bf{a}}(r,\theta )}}{{\partial \theta }}} \right\|^2}},\ $and $\footnotesize {c \triangleq \frac{{\partial {{\bf{a}}^H}(r,\theta )}}{{\partial \theta }}{\bf{a}}(r,\theta )\ }$are relevant intermediate parameters. Similarly, other relevant terms in (\ref{theta}) and (\ref{dis}) can be obtained as
\begin{equation}
	\setlength\abovedisplayskip{1.5pt}
	\setlength\belowdisplayskip{1.5pt}
%	\hspace{-1.7ex}
\footnotesize
		\begin{aligned}
		{\| {{{\bf{g}}_r}} \|^2} &= 2Mp + 2{| q |^2},~\Re \{ {{{\bf{g}}^H}{\bf{Wg}}} \} = 2M^2\Re\{Me-c^*q\},\\
		{\sin ^2}\Omega  &= \frac{{Ma - {{| c |}^2}}}{{Ma + {{| c |}^2}}},~{\sin ^2}\Theta  = \frac{{Mp - {{| q |}^2}}}{{Mp + {{| q |}^2}}},
		\label{parameter2}
		\end{aligned}
\end{equation}
where the intermediate parameters $e$, $p$, and $q$ are defined as
\begin{equation}
	\setlength\abovedisplayskip{1.5pt}
	\setlength\belowdisplayskip{1.5pt}
\footnotesize
	\begin{aligned}
		e \triangleq \frac{{\partial {{\bf{a}}^H}(r,\theta )}}{{\partial \theta }}\frac{{\partial {\bf{a}}(r,\theta )}}{{\partial r}},p \triangleq {\left\| {\frac{{\partial {\bf{a}}(r,\theta )}}{{\partial r}}} \right\|^2},q \triangleq \frac{{\partial {{\bf{a}}^H}(r,\theta )}}{{\partial r}}{\bf{a}}(r,\theta ).
		\label{8}
	\end{aligned}
\end{equation}
Note that all these intermediate parameters are dependent on ${\bf{a}}(r,\theta )$ and its derivatives. Therefore, to further derive the expression of intermediate parameters, based on (\ref{eq2}), the derivatives of $a_m(r,\theta)$ with respect to $r$ and $\theta$ are obtained as
\begin{equation}
\setlength\abovedisplayskip{1.5pt}
\setlength\belowdisplayskip{1pt}
%\hspace{-2.8ex}
 \scriptsize
\begin{aligned}
\frac{{\partial {a_m}(r,\theta )}}{{\partial \theta }}= -{ \frac{{j2\pi }}{\lambda }}{e^{ - j\frac{{2\pi }}{\lambda }{r_m}}}\frac{{\partial {r_m}}}{{\partial \theta }},~
\frac{{\partial {a_m}(r,\theta )}}{{\partial r}}= -{ \frac{{j2\pi }}{\lambda }}{e^{ - j\frac{{2\pi }}{\lambda }{r_m}}}\frac{{\partial {r_m}}}{{\partial r}},	
\label{derivative1}
\end{aligned}
\end{equation}
where $\frac{{\partial {r_m}}}{{\partial \theta }} $ and $\frac{{\partial {r_m}}}{{\partial r}}$ can be obtained based on (\ref{eq9}):
\begin{equation}
\setlength\abovedisplayskip{2pt}
\setlength\belowdisplayskip{2pt}
\footnotesize
\begin{aligned}
\frac{{\partial {r_m}}}{{\partial \theta }} &= \frac{{ - m{d_T}\cos \theta }}{{\sqrt {1 - 2m{\varepsilon _T}\sin \theta  + {{(m{\varepsilon _T})}^2}} }},\\
\frac{{\partial {r_m}}}{{\partial r}}& = \frac{{1 - m{\varepsilon _T}\sin \theta }}{{\sqrt {1 - 2m{\varepsilon _T}\sin \theta  + {{(m{\varepsilon _T})}^2}} }}.
\end{aligned}
\label{derivative2}
\end{equation}
%\begin{equation}
%	\setlength\abovedisplayskip{1pt}
%	\setlength\belowdisplayskip{1pt}
%	\footnotesize
%	\begin{aligned}
%		\frac{{\partial {r_m}}}{{\partial \theta }} &= \frac{\partial }{{\partial \theta }}\left( {r\sqrt {1 - 2m{\varepsilon _T}\sin \theta  + {{(m{\varepsilon _T})}^2}} }\right) \\
%		&= \frac{{ - m{d_T}\cos \theta }}{{\sqrt {1 - 2m{\varepsilon _T}\sin \theta  + {{(m{\varepsilon _T})}^2}} }},\\
%		\frac{{\partial {r_m}}}{{\partial r}} &= \frac{\partial }{{\partial r}}\left( {r\sqrt {1 - 2m{\varepsilon _T}\sin \theta  + {{(m{\varepsilon _T})}^2}} }\right )\\
%		&= \frac{{1 - m{\varepsilon _T}\sin \theta }}{{\sqrt {1 - 2m{\varepsilon _T}\sin \theta  + {{(m{\varepsilon _T})}^2}} }}.
%	\end{aligned}
%	\label{derivative2}
%\end{equation}
As a result, the intermediate parameters $a$, $c$, $e$, $q$, and $q$ are given by
\begin{equation}
	\setlength\abovedisplayskip{1.5pt}
	\setlength\belowdisplayskip{2pt}
	\footnotesize
	\hspace{3.1ex}
	\begin{aligned}
		%		\begin{array}{l}
			a &= {\sum\nolimits_{m =  - \frac{{M = 1}}{2}}^{\frac{{M = 1}}{2}} {\left( {\frac{{\partial {a_m}(r,\theta )}}{{\partial \theta }}} \right)} ^2}\\
			&= \frac{{4{\pi ^2}{r^2}{{\cos }^2}\theta }}{{{\lambda ^2}}}\sum\nolimits_{m =  - \frac{{M - 1}}{2}}^{\frac{{M - 1}}{2}} {\frac{{{m^2}{\varepsilon _T}^2}}{{1 - 2m{\varepsilon _T}\sin \theta  + {{(m{\varepsilon _T})}^2}}}},
			%		\end{array}\\
	\label{psuma}
	\end{aligned}
\end{equation}
\begin{equation}
	\setlength\abovedisplayskip{1pt}
	\setlength\belowdisplayskip{1pt}
\footnotesize
%		\hspace{-2.6ex}
		\begin{aligned}
			%			\begin{array}{l}
				c &=\sum\nolimits_{m =  - \frac{{M = 1}}{2}}^{\frac{{M = 1}}{2}} {\frac{{\partial {a_m}^*(r,\theta )}}{{\partial \theta }}{a_m}(r,\theta )}\\ 
				&=  - j\frac{{2\pi r\cos \theta }}{\lambda }\sum\nolimits_{m =  - \frac{{M - 1}}{2}}^{\frac{{M - 1}}{2}} {\frac{{m{\varepsilon _T}}}{{\sqrt {1 - 2m{\varepsilon _T}\sin \theta  + {{(m{\varepsilon _T})}^2}} }}},
		\end{aligned}
		\label{psumc}
\end{equation}
\begin{equation}
	\setlength\abovedisplayskip{1pt}
	\setlength\belowdisplayskip{2pt}
\footnotesize
			\hspace{-13.6ex}
		\begin{aligned}
			e &= \sum\nolimits_{m =  - \frac{{M = 1}}{2}}^{\frac{{M = 1}}{2}} {\frac{{\partial {a_m}^*(r,\theta )}}{{\partial \theta }}\frac{{\partial {a_m}(r,\theta )}}{{\partial r}}} \\
			&= \frac{{4{\pi ^2}r}}{{{\lambda ^2}}}\sum\nolimits_{m =  - \frac{{M - 1}}{2}}^{\frac{{M - 1}}{2}} {\frac{{m{\varepsilon _T}\cos \theta (m{\varepsilon _T}\sin \theta  - 1)}}{{1 - 2m{\varepsilon _T}\sin \theta  + {{(m{\varepsilon _T})}^2}}}},
		\end{aligned}
		\label{psume}
\end{equation}
\begin{equation}
	\setlength\belowdisplayskip{2pt}
	\footnotesize
			\hspace{-8ex}
		\begin{aligned}
			p &= {\sum\nolimits_{m =  - \frac{{M = 1}}{2}}^{\frac{{M = 1}}{2}} {\left( {\frac{{\partial {a_m}(r,\theta )}}{{\partial r}}} \right)} ^2}\\
			&= \frac{{4{\pi ^2}}}{{{\lambda ^2}}}\sum\nolimits_{m =  - \frac{{M - 1}}{2}}^{\frac{{M - 1}}{2}} {\left[ {1 - \frac{{{m^2}{\varepsilon _T}^2{{\cos }^2}\theta }}{{1 - 2m{\varepsilon _T}\sin \theta  + {{(m{\varepsilon _T})}^2}}}}\right ]},
		\end{aligned}
		\label{psump}
\end{equation}
\begin{equation}
	\setlength\abovedisplayskip{1pt}
	\setlength\belowdisplayskip{2pt}
\footnotesize
			\hspace{-12ex}
		\begin{aligned}
			q &= \sum\nolimits_{m =  - \frac{{M = 1}}{2}}^{\frac{{M = 1}}{2}} {\frac{{\partial {a_m}^*(r,\theta )}}{{\partial r}}{a_m}(r,\theta )} \\
			&= j\frac{{2\pi }}{\lambda }\sum\nolimits_{m =  - \frac{{M - 1}}{2}}^{\frac{{M - 1}}{2}} {\frac{{1 - m{\varepsilon _T}\sin \theta }}{{\sqrt {1 - 2m{\varepsilon _T}\sin \theta  + {{(m{\varepsilon _T})}^2}} }}}.
		\end{aligned}
		\label{psumq}
\end{equation}
Since$\ {\varepsilon _T} \ll 1,\ $similar to\cite{b2}, we can derive the closed-form expressions for the above intermediate parameters as follows.
\begin{prop}\label{prop1}
The closed-form expressions of the intermediate parameters $a,c,e,p,q$ can be derived as
\begin{equation}
	\setlength\abovedisplayskip{1pt}
	\setlength\belowdisplayskip{2pt}
\footnotesize
\hspace{-9.8ex}
	\begin{aligned}
%		\begin{array}{l}
	a &= \frac{{4{\pi ^2}{r^2}{{\cos }^2}\theta }}{{{\lambda ^2}{\varepsilon _T}}}\left[ {\frac{{{D_T}}}{r} + \sin \theta \ln \left| {\frac{{\frac{{D_T^2}}{{4{r^2}}} - \sin \theta\frac{{{D_T} }}{r} + 1}}{\frac{{D_T^2}}{{4{r^2}}} + \sin \theta\frac{{{D_T} }}{r} + 1}} \right|} \right.\\
	&\left. { - \frac{{\cos 2\theta }}{{\cos \theta }}\Delta _{{\rm{span}}}^{\rm{t}}\left(\frac{{{D_T}}}{r}\right)} \right ],
%		\end{array}\\
	\label{parametera}
	\end{aligned}
\end{equation}
\begin{equation}
	\setlength\abovedisplayskip{1pt}
	\setlength\belowdisplayskip{2pt}
\footnotesize
	\hspace{1.6ex}
	\begin{aligned}
%			\begin{array}{l}
	c &=  - j\frac{{2\pi r\cos \theta }}{\lambda }\left[ { {\sqrt {\frac{{D_T^2}}{{4{r^2}}} - \sin \theta\frac{{{D_T} }}{r} + 1} } } 
	 { - \sqrt {\frac{{D_T^2}}{{4{r^2}}} + \sin \theta\frac{{{D_T} }}{r} + 1} } \right.\\ 
	&\left.+  \psi \left(\frac{D_T}{r}\right)\sin \theta \right],
%		\end{array}\\
	\end{aligned}
	\label{parameterc}
\end{equation}
\begin{equation}
	\setlength\abovedisplayskip{1pt}
	\setlength\belowdisplayskip{2pt}
\footnotesize
%	\hspace{-3.8ex}
		\begin{aligned}
	e & = \frac{{4{\pi ^2}r\cos \theta }}{{{\lambda ^2}{\varepsilon _T}}}\left[\frac{{{D_T}}}{r}\sin\theta - \frac{{\cos 2\theta }}{2}\ln \left| {\frac{{\frac{{D_T^2}}{{4{r^2}}} - \sin \theta\frac{{{D_T} }}{r} + 1}}{{\frac{{D_T^2}}{{4{r^2}}} + \sin \theta\frac{{{D_T} }}{r} + 1}}} \right|\right.\\
	&\left.{ - \Delta _{{\rm{span}}}^{\rm{t}}\left(\frac{{{D_T}}}{r}\right)}\sin 2\theta \right ],
		\end{aligned}
	\label{parametere}
\end{equation}
\begin{equation}
	\setlength\belowdisplayskip{2pt}
\footnotesize
%	\hspace{-3.8ex}
		\begin{aligned}
p &= \frac{{4{\pi ^2}}}{{{\lambda ^2}{\varepsilon _T}}}\left[ {{{\sin }^2}\theta \frac{{{D_T}}}{r} - \ln \left| {\frac{{\frac{{D_T^2}}{{4{r^2}}} - \sin \theta\frac{{{D_T} }}{r} + 1}}{{\frac{{D_T^2}}{{4{r^2}}} + \sin \theta\frac{{{D_T} }}{r} + 1}}} \right|} {{\cos }^2}\theta \sin \theta \right.\\
&\left. { + \Delta _{{\rm{span}}}^{\rm{t}}\left(\frac{{{D_T}}}{r}\right)}\cos \theta \cos 2\theta  \right],
		\end{aligned}
	\label{parameterp}
\end{equation}
\begin{equation}
	\setlength\abovedisplayskip{1pt}
	\setlength\belowdisplayskip{2pt}
\footnotesize
%	\hspace{-3.8ex}
	\begin{aligned}
q &= j\frac{{2\pi }}{{\lambda {\varepsilon _T}}}\left[  \psi\left(\frac{{{D_T}}}{r}\right){{{\cos }^2}\theta - \sin \theta  {\sqrt {\frac{{D_T^2}}{{4{r^2}}} - \sin \theta\frac{{{D_T} }}{r} + 1} } } \right.\\
&\left.{ { +\sin \theta {\sqrt {\frac{{D_T^2}}{{4{r^2}}} + \sin \theta\frac{{{D_T} }}{r} + 1} } } } \right],
	\end{aligned}
	\label{parameterq}
\end{equation}
where$\ \Delta _{{\rm{span}}}^{\rm{t}}(\frac{{{D_T}}}{r}) \triangleq \arctan ( {\frac{{{D_T}}}{{{\rm{2}}r\cos \theta }} - \tan \theta } ) + \arctan ( {\frac{{{D_T}}}{{{\rm{2}}r\cos \theta }} + \tan \theta } )\ $is the angular span of the transmit array\cite{b2}, $\psi(\frac{{{D_T}}}{r})\triangleq \ln \left(\frac{{{p_2} + \sqrt {1 + p_2^2} }}{{{p_1} + \sqrt {1 + p_1^2} }}\right)$ and$\ {p_1} = \frac{{ - \frac{{{D_T}}}{{2r}} - \sin \theta }}{{\cos \theta }},~{p_2} = \frac{{\frac{{{D_T}}}{{2r}} - \sin \theta }}{{\cos \theta }}.\ $
\begin{IEEEproof}
	Please refer to Appendix A.
\end{IEEEproof}
\end{prop}
Based on (\ref{parametera})-(\ref{parameterq}), the parameters in (\ref{parameter1}) and (\ref{parameter2}) can be obtained. By substituting them into (\ref{theta}) and (\ref{dis}), the closed-form expressions of the near-field CRBs of angle and range can be obtained, as given below.
\begin{theorem}\label{theorem1}
For near-field monostatic XL-MIMO radar mode, the CRBs of angle $\theta$ and range $r$ can be expressed in closed-form as
\begin{equation}
\setlength\abovedisplayskip{1pt}
\setlength\belowdisplayskip{-1pt}	
\footnotesize
%\hspace{-1.25ex}
\begin{aligned}
{CR{B_\theta } = \frac{1}{{2{\gamma }L}}\frac{M{( {Mp - {{| q |}^2}} )}}{{2\{ {( {Ma - {{| c |}^2}} )( {Mp - {{| q |}^2}} ) - {{[ {Me - {c^*}q} ]}^2}} \}}}},
\label{crb-mimo1}
\end{aligned}
\end{equation}
\begin{equation}
\setlength\abovedisplayskip{-1pt}
\setlength\belowdisplayskip{1pt}
\footnotesize
%\hspace{-1.2ex}
\begin{aligned}
{CR{B_r} = \frac{1}{{2\gamma L}}\frac{M{( {Ma - {{| c |}^2}} )}}{{2\{ {( {Ma - {{| c |}^2}} )( {Mp - {{| q |}^2}} ) - {{[ {Me - {c^*}q} ]}^2}} \}}}},
\label{crb-mimo2}
\end{aligned}
\end{equation}
where$\ \gamma  = \frac{{P{{| \kappa  |}^2}}}{{{\sigma ^2}}}\ $defined as the SNR, with $\sigma^2=N_0B$, and  $L\triangleq BT_p$ denotes time-bandwidth product in a CPI. The intermediate parameters $a,c,e,p,q$ are given in closed-form in Proposition \ref{prop1}.

To gain useful insights for Theorem \ref{theorem1}, we consider three asymptotic cases when$\ \frac{{{D_T}}}{r} \gg 1,\ $$\frac{D_T}{r\cos\theta}\to \infty$, or$\ \frac{{{D_T}}}{r} \ll 1,\ $respectively. 
\end{theorem}

\begin{corollary}{\label{t1}}
When$\ \frac{{{D_T}}}{r} \gg 1,\ $the CRBs in Theorem \ref{theorem1} reduces to
\begin{equation}
	\setlength\abovedisplayskip{1.5pt}
	\setlength\belowdisplayskip{1pt}
\footnotesize
	\hspace{-3.3ex}
	\begin{aligned}
	&CR{B_\theta } = \frac{1}{{2\gamma L}}\times\\
& \frac{{{\lambda ^2}\left[ { {{\left( {\frac{{{D_T}\sin\theta}}{r}} \right)}^2} + \frac{{\pi{D_T}}}{r}\cos \theta \cos 2\theta  - 4{{( {{{\cos }^2}\theta \ln \frac{{{D_T}}}{{r\cos \theta }} + {{\sin }^2}\theta } )}^2}} \right]}}{{8{\pi ^2}{r^2}M ( {\frac{{{\pi D_T}}}{r{\cos \theta }} - 4{{\ln }^2}\frac{{{D_T}}}{{r\cos \theta }}} ){{\cos }^2}\theta}},
	\end{aligned}
\label{extt}
\end{equation}
\begin{equation}
	\setlength\abovedisplayskip{1pt}
	\setlength\belowdisplayskip{1pt}
\footnotesize
%	\hspace{-2ex}
	\begin{aligned}
		&CR{B_r} =\frac{1}{{2\gamma L}}\frac{{{\lambda ^2}\left[ {{{\left( {\frac{{{D_T}}}{r}} \right)}^2} + \frac{{{\pi D_T}}}{r}\frac{{\cos 2\theta }}{{\cos \theta }}  - 4 {{( {\ln \frac{{{D_T}}}{{r\cos \theta }} - 1} )}^2}}{{\sin }^2}\theta \right]}}{{8{\pi ^2}M( {\frac{{\pi {D_T}}}{{r\cos \theta }} - 4{{\ln }^2}\frac{{{D_T}}}{{r\cos \theta }}} )}}.
	\end{aligned}
	\label{extr}
\end{equation}
\begin{IEEEproof}
	Please refer to Appendix B.
\end{IEEEproof}
\end{corollary}
\begin{corollary}\label{t2}
When $\frac{D_T}{r\cos\theta}\to \infty$, the limits of near-field CRBs of angle and range are
\begin{equation}
\setlength\abovedisplayskip{1pt}
\setlength\belowdisplayskip{1pt}
\footnotesize
\begin{aligned}
 \;\mathop {\lim }\limits_{\frac{{{D_T}}}{{r\cos \theta }} \to \infty } CR{B_\theta } = \frac{1}{{2\gamma L}}\frac{{{\lambda ^2}{d_T}{{\sin }^2}\theta }}{{8{\pi ^3}{r^3}\cos \theta }},\
			\label{crb-mimo-inf1}
\end{aligned}
\end{equation}
\begin{equation}
\setlength\abovedisplayskip{1pt}
\setlength\belowdisplayskip{-2pt}
\footnotesize
\begin{aligned}
 \;\mathop {\lim }\limits_{\frac{{{D_T}}}{{r\cos \theta }} \to \infty } CR{B_r} = \frac{1}{{2\gamma L}}\frac{{{\lambda ^2}{d_T}\cos \theta }}{{8{\pi ^3}r}}.\
	\label{crb-mimo-inf2}
\end{aligned}
\end{equation}
\begin{IEEEproof}
	Please refer to Appendix C.
\end{IEEEproof}
\end{corollary}
%The above result shows that when the transmit array aperture $D_T$ is much larger than the perpendicular range $r\cos\theta$, both the CRB of the angle and range are independent of the array aperture. Besides, they are proportional to the element-spacing $d_T$, which means even if the array aperture is infinity, the parameter estimation error can not reduce to zero. However, they can be further reduced by decreasing the element-spacing.
As a comparison, the CRB for angle with the conventional far-field UPW model can be derived based on\cite{b28}
\begin{equation}
\setlength\abovedisplayskip{0pt}
\setlength\belowdisplayskip{1pt}
\footnotesize
\begin{aligned}
C_{\theta} &= \frac{1}{{2\gamma L}}\frac{{3{\lambda ^2}}}{{2{\pi ^2}d_T^2M ( {{M^2} - 1} ){{\cos }^2}\theta}},
	\label{far-field}
\end{aligned}
\end{equation}
whereas that for range is infinity since far-field sensing does not have range discrimination capability. 
Note that for the given inter-element distance $d_T$, the array aperture $D_T$ increases proportionally with the number of antennas $M$. Therefore, it is observed from (\ref{crb-mimo-inf1})-(\ref{far-field}) that different from the far-field UPW model where the CRB decreases indefinitely with the number of antennas $M$ or array aperture $D_T$, for near-field XL-MIMO sensing with USW model, when the array aperture $D_T$ goes to infinity, the CRBs of angle and range approach to a limit that is dependent on the inter-element distance $d_T$. This implies that in order to achieve the extreme performance for unbiased angle and range estimation with zero errors, not only the array aperture $D_T$ should be sufficiently large, but also the inter-element spacing $d_T$ should be infinitesimally small. This theoretically shows that the emerging holographic MIMO could potentially achieve better near-field sensing than conventional discrete MIMO \cite{b77,b78,b79}.
% Furthermore, the CRB for range in far-field model is infinity since there is no range estimation ability.
\begin{corollary}\label{jinsifar}
When $\frac{D_T}{r}\ll 1$, the CRB in (\ref{crb-mimo1}) reduces to
\begin{equation}
	\setlength\abovedisplayskip{0pt}
	\setlength\belowdisplayskip{1pt}
	\footnotesize
	\begin{aligned}
		CRB_{\theta} = \frac{1}{{2\gamma L}}\frac{{3{\lambda ^2}}}{{2{\pi ^2}d_T^2M^3{{\cos }^2}\theta }}\Xi (\theta ),
		\label{jinsifar1}
	\end{aligned}
\end{equation}
where$\ \Xi (\theta ) \triangleq \frac{{6{{\sin }^2}\theta  + {{\cos }^2}\theta \cos 2\theta }}{{9{{\sin }^2}\theta  + {{\cos }^6}\theta }} \in (0.6,1].\ $
\begin{IEEEproof}
	Please refer to Appendix D.
\end{IEEEproof}
\end{corollary}
By comparing (\ref{far-field}) and (\ref{jinsifar1}), it is found that when the array aperture $D_T$ is much smaller than the target range $r$, our newly derived CRB for USW-based near-field sensing reduces to the conventional UPW-based far-field sensing, with a correction factor $\Xi (\theta )$.

The near-field CRB expressions in (\ref{crb-mimo1}) and (\ref{crb-mimo2}) are derived by considering the USW propagation based on the exact distance expression (\ref{eq9}). Another common approach for modelling USW is to apply the second-order Taylor approximation for (\ref{eq9})\cite{b33}\cite{b36}. In this case, the CRBs in (\ref{crb-mimo1})-(\ref{crb-mimo2}) can be obtained in the following.    
\begin{corollary}{\label{lem4}}
When the second-order Taylor approximation is used for the distance expression (\ref{eq9}), the near-field CRBs in (\ref{crb-mimo1})-(\ref{crb-mimo2}) reduce to
%\begin{align}	\label{crb-taylor-mimo1}
%	\footnotesize
%	CR{B_\theta } &= \frac{1}{{2{T_p}\gamma }}\frac{{3{\lambda ^2}}}{{2{\pi ^2}d_T^2{{\cos }^2}\theta M( {{M^2} - 1} )}},\\\label{crb-taylor-mimo2}
%		CR{B_r} &= \frac{1}{{2{T_p}\gamma }}\frac{{6{\lambda ^2}{r^2}[ {15{r^2} + {{( {{d_T}{\rm{sin}}\theta } )}^2}( {{M^2} - 4} )} ]}}{{{{( {\pi {d^2}_T{{\cos }^2}\theta } )}^2}M( {{M^2} - 1} )( {{M^2} - 4} )}}.
%\end{align} 
\begin{equation}
\setlength\abovedisplayskip{1pt}
\setlength\belowdisplayskip{1pt}
\hspace{-8.5ex}
	\footnotesize
	\begin{aligned}
		CR{B_\theta } = \frac{1}{{2\gamma L}}\frac{{3{\lambda ^2}}}{{2{\pi ^2}d_T^2M ( {{M^2} - 1} ){{\cos }^2}\theta}},
	\label{crb-taylor-mimo1}
	\end{aligned}
\end{equation}
\begin{equation}
	\setlength\abovedisplayskip{1pt}
	\setlength\belowdisplayskip{1pt}
	\footnotesize
	\begin{aligned}
		CR{B_r} = \frac{1}{{2\gamma L}}\frac{{6{\lambda ^2}{r^2}[ {15{r^2} + {{( {{d_T}{\rm{sin}}\theta } )}^2}( {{M^2} - 4} )} ]}}{{{{( {\pi d^2_T{{\cos }^2}\theta } )}^2}M( {{M^2} - 1} )( {{M^2} - 4} )}}.
		\label{crb-taylor-mimo2}
	\end{aligned}
\end{equation}
	\begin{IEEEproof}
		Please refer to Appendix E.
	\end{IEEEproof}
\end{corollary}
Corollary \ref{lem4} shows that surprisingly, if the second-order Taylor approximation is used for USW modeling, the CRB of angle coincides with that based on the conventional UPW model in (\ref{far-field}). This is because the USW feature, which is mainly captured by the second-order term in ${\bf{a}}(r,\theta ),\ $is offset by other terms $p,e,c,q$. This implies that the second-order Taylor approximation may not be accurate enough to evaluate the near-field CRB for angle estimation. On the other hand, when $M$ is large, the expression (\ref{crb-taylor-mimo2}) shows that the CRB of range is inversely proportional to $M^3$ and would eventually approach to 0. This is in contrast with the result based on the exact distance expression as presented in (\ref{crb-mimo-inf2}).
%	\begin{lemma}
	%		When the target is located at the x-axis, i.e., $\theta=0$, (\ref{crb-taylor-mimo}) can be reduced to
	%		\begin{equation}
		%			\ \begin{array}{*{20}{l}}
			%				{CR{B_\theta } = \frac{1}{{2\gamma }}\frac{6}{{{\pi ^2}( {{M^2} - 1} )}}}\\
			%				{CR{B_r} = \frac{1}{{2\gamma }}\frac{{360{r^4}}}{{{\pi ^2}{d^2}( {{M^2} - 1} )( {{M^2} - 4} )}}}
			%			\end{array}\
		%		\label{lem5}
		%		\end{equation}
	%	\end{lemma}
\subsection{XL-phased Array Radar}
For monostatic XL-phased array radar, we have$\ {\bf{g}} = {\bf{a}}(r,\theta ),$ and$\ \rho  = \kappa \sqrt {{T_p}PM}\ $for the model (\ref{model}). Thus,$\ {\| {\bf{g}} \|^2} = {\| {{\bf{a}}(r,\theta )} \|^2} = M.\ $Similar to the analysis of XL-MIMO radar in Section \ref{monomimo}, the following parameters in (\ref{theta}) and (\ref{dis}) can be obtained:
\begin{equation}
		\setlength\abovedisplayskip{2pt}
	\setlength\belowdisplayskip{2pt}
\footnotesize
	\begin{aligned}
		%	\ \begin{array}{l}
			{\| {{{\bf{g}}_\theta }} \|^2} = a,
			{\| {{{\bf{g}}_r}} \|^2} = p,{\sin ^2}\Omega  = 1 - \frac{{{{| c |}^2}}}{{aM}},{\sin ^2}\Theta  = 1 - \frac{{{{| q |}^2}}}{{pM}},
			%	\end{array}\
	\end{aligned}
\end{equation}
where the intermediate parameters $a$, $c$, $p$, and $q$ are given in closed-form in (\ref{parametera})-(\ref{parameterq}). Therefore, based on (\ref{theta}) and (\ref{dis}), the closed-form CRBs of range and angle can be obtained, as shown in Theorem \ref{theoremph} below.
%Furthermore, the determinant of the closed-form expression of CRBs in (\ref{theta}) and (\ref{dis}) ${\bf{Q}}'$ can be expressed as
%\begin{equation}
%	\setlength\abovedisplayskip{1.5pt}
%	\setlength\belowdisplayskip{1.5pt}
%	\footnotesize
%	\begin{aligned}
%		\ \det {{\bf{Q}}^\prime } = \frac{{( {Ma - {{| c |}^2}} )( {Mp - {{| q |}^2}} ) - {{( {Me - {c^*}q} )}^2}}}{{{M^2}}}.\
%	\end{aligned}
%\end{equation}
\begin{theorem}\label{theoremph}
For near-field monostatic XL-phased array radar sensing, the closed-form CRBs for angle and range estimation are
\begin{equation}
\setlength\abovedisplayskip{2pt}
\setlength\belowdisplayskip{2pt}
\footnotesize
\begin{aligned}
	{CR{B_\theta } = \frac{1}{{2\gamma L}}\frac{{( {Mp - {{| q |}^2}} )}}{{ {( {Ma - {{| c |}^2}} )( {Mp - {{| q |}^2}} ) - {{[ {Me - {c^*}q} ]}^2}} }}},
\end{aligned}
\label{crb-ph1}
\end{equation}
\begin{equation}
	\setlength\abovedisplayskip{2pt}
	\setlength\belowdisplayskip{2pt}
\footnotesize
	\begin{aligned}
	{CR{B_r} = \frac{1}{{2\gamma L}}\frac{{( {Ma - {{| c |}^2}} )}}{{ {( {Ma - {{| c |}^2}} )( {Mp - {{| q |}^2}} ) - {{[ {Me - {c^*}q} ]}^2}} }}},
\end{aligned}
\label{crb-ph2}
\end{equation}
where the closed-form expression of the intermediate parameters $a,c,e,p,q$ are given in (\ref{parametera})-(\ref{parameterq}). 

By comparing Theorem \ref{theoremph} with Theorem \ref{theorem1}, it is observed that the CRBs for XL-phased array radar mode are $\frac{2}{M}$ fractional of those for XL-MIMO radar mode. Such an additional gain is contributed by the transmit beamforming gain by phased array radar. Similar analysis as in Section \ref{monomimo} can be obtained for XL-phased array sensing. The proofs are omitted for brevity.
\end{theorem}
\begin{corollary}\label{ph-inf}
For the asymptotic case when$\ \frac{{{D_T}}}{r} \gg 1\ $, the CRBs of angle and range in Theorem \ref{theoremph} reduce to
\begin{equation}
	\setlength\abovedisplayskip{1.5pt}
	\setlength\belowdisplayskip{1.5pt}
\footnotesize
	\hspace{-1ex}
	\begin{aligned}
		&CR{B_\theta } \approx \frac{1}{{2\gamma L}}\times\\
		& \frac{{{\lambda ^2}\left[ { {{\left( {\frac{{{D_T}}}{r}} \right)}^2}{{\sin }^2}\theta +   \frac{{\pi{D_T}}}{r} \cos \theta \cos 2\theta- 4{{( {{{\cos }^2}\theta \ln \frac{{{D_T}}}{{r\cos \theta }} + {{\sin }^2}\theta } )}^2}} \right]}}{{4{\pi ^2}{r^2}M^2 ( {\frac{{{\pi D_T}}}{r{\cos \theta }} - 4{{\ln }^2}\frac{{{D_T}}}{{r\cos \theta }}}){{\cos }^2}\theta }},\\
	\end{aligned}
	\label{extt2}
\end{equation}
\begin{equation}
	\setlength\abovedisplayskip{1.5pt}
	\setlength\belowdisplayskip{1.5pt}
\footnotesize
	\hspace{1ex}
	\begin{aligned}
		&CR{B_r} \approx \frac{1}{{2\gamma L}} \frac{{{\lambda ^2}\left[ {{{\left( {\frac{{{D_T}}}{r}} \right)}^2} + \frac{\pi{{D_T}}}{r}\frac{{\cos 2\theta }}{{\cos \theta }}  - 4 {{( {\ln \frac{{{D_T}}}{{r\cos \theta }} - 1} )}^2}}{{\sin }^2}\theta \right]}}{{4{\pi ^2}M^2 ({\frac{{\pi {D_T}}}{{r\cos \theta }} - 4{{\ln }^2}\frac{{{D_T}}}{{r\cos \theta }}} )}}.
	\end{aligned}
	\label{extr2}
\end{equation}
\begin{corollary}\label{co1}
For the asymptotic case when $\frac{D_T}{r\cos\theta}\to \infty$, the CRBs in (\ref{extt2}) and (\ref{extr2}) approach to the following limits:
\begin{equation}
	\setlength\abovedisplayskip{2pt}
\setlength\belowdisplayskip{2pt}
\footnotesize
	\begin{aligned}
	 \;\mathop {\lim }\limits_{\frac{{{D_T}}}{{r\cos \theta }} \to \infty } CR{B_\theta }= \frac{1}{{2\gamma L}}\frac{{{\lambda ^2}{d_T}{{\sin }^2}\theta }}{{4M{\pi ^3}{r^3}\cos \theta }},
\label{inf-ph1}
	\end{aligned}
\end{equation}
\begin{equation}
	\setlength\abovedisplayskip{0pt}
\setlength\belowdisplayskip{0pt}
\footnotesize
	\hspace{-3ex}
	\begin{aligned}
 \;\mathop {\lim }\limits_{\frac{{{D_T}}}{{r\cos \theta }} \to \infty } CR{B_r}= \frac{1}{{2\gamma L}}\frac{{{\lambda ^2}{d_T}\cos \theta }}{{4M{\pi ^3}r}}.
\label{inf-ph2}
	\end{aligned}
\end{equation}
\end{corollary}
Corollary \ref{co1} shows that different from the near-field XL-MIMO radar mode, for XL-phased array radar mode, the CRBs decrease indefinitely as the number of array elements $M$ or aperture $D_T$ go to sufficiently large.
\end{corollary}

The CRB of angle for the conventional far-field UPW-based sensing can be obtained based on \cite{b67}
\begin{equation}
	\footnotesize
	\setlength\abovedisplayskip{1pt}
	\setlength\belowdisplayskip{1.5pt}
	\begin{aligned}
		\begin{aligned}
			{C_\theta } = \frac{1}{{2L{\gamma}}}\frac{{3{\lambda ^2}}}{{{\pi ^2}d_T^2M^2( {{M^2} - 1} ){{\cos }^2}\theta }}.\
			\label{upw-ph}
		\end{aligned}
	\end{aligned}
\end{equation}
\begin{corollary}\label{jinsifar3}
Similar to (\ref{jinsifar1}), when $\frac{D_T}{r}\ll 1$, the CRB in (\ref{crb-ph1}) reduces to
	\begin{equation}
		\setlength\abovedisplayskip{0pt}
		\setlength\belowdisplayskip{1pt}
		\footnotesize
		\begin{aligned}
			CRB_{\theta} = \frac{1}{{2\gamma L}}\frac{{3{\lambda ^2}}}{{{\pi ^2}d_T^2M^4{{\cos }^2}\theta }}\Xi (\theta ),
			\label{jinsifar2}
		\end{aligned}
	\end{equation}
where$\ \Xi (\theta ) = \frac{{6{{\sin }^2}\theta  + {{\cos }^2}\theta \cos 2\theta }}{{9{{\sin }^2}\theta  + {{\cos }^6}\theta }}.\ $
\end{corollary}
\begin{corollary}
When the second-order Taylor approximation is used for the distance expression (\ref{eq9}), the CRBs in (\ref{crb-ph1}) and (\ref{crb-ph2}) reduce to 
\begin{equation}
		\footnotesize
		\setlength\abovedisplayskip{1pt}
		\setlength\belowdisplayskip{1pt}
			\hspace{-16ex}
		\begin{aligned}
			CR{B_\theta } &= \frac{1}{{2\gamma L}}\frac{{3{\lambda ^2}}}{{{\pi ^2}d_T^2M^2( {{M^2} - 1} ){{\cos }^2}\theta }},
		\end{aligned}
\label{crb-taylor-ph1}
\end{equation}
\begin{equation}
	\footnotesize
	\setlength\abovedisplayskip{0pt}
	\setlength\belowdisplayskip{1pt}
	\hspace{6ex}
	\begin{aligned}
			CR{B_r} & = \frac{1}{{2\gamma L}}\frac{{12{\lambda ^2}{r^2}[ {15{r^2} + {{( {{d_T}{\rm{sin}}\theta } )}^2}( {{M^2} - 4} )} ]}}{{{{( {\pi {d^2}_T{{\cos }^2}\theta } )}^2}M^2( {{M^2} - 1} )( {{M^2} - 4} )}},
		\end{aligned}
		\label{crb-taylor-ph2}
\end{equation}
\end{corollary}
which are consistent with the results in \cite{b41}.

%\begin{lemma}
%	when $\theta=0$, the CRBs of XL-phased array radar in (\ref{crb-taylor-ph1})-(\ref{crb-taylor-ph2}) can be reduced to
%	\begin{equation}
%	\ \begin{array}{*{20}{l}}
%		{CR{B_\theta } = \frac{1}{{2\gamma }}\frac{{3{\lambda ^2}}}{{{\pi ^2}d_T^2( {{M^2} - 1} )}}}\\
%		{CR{B_r} = \frac{1}{{2\gamma }}\frac{{720{r^4}}}{{{\pi ^2}{\lambda ^2}({M^2} - 4)({M^2} - 1)}}}
%	\end{array}\
%	\end{equation}
%\end{lemma}
\section{Near-Field CRB for Bistatic Sensing}\label{bistaticsensing}
%In this section, as illustrated in Fig.\ref{bi}, we consider a bistatic XL-sensing system when one of the range from transmitter to target or from target to receiver is large, which means the transmit or the receive antenna array can be viewed as a small array and modeled based on far-field assumption. Such kind of assumption is practical and useful, since for MIMO radar, forming virtual array is effective to achieve higher angle resolution\cite{b27}. In particular, when the sparsity of transmit/receive array is exploited, an virtural array with $MN$ elements using only $M+N$ array elements can be obtained.
In this section, we extend the above analysis to bistatic near-field sensing. However, when near-field USW model is considered at both the transmitter and receiver sides, the CRB derivation is rather challenging. Moreover, considering the practical transmitter and receiver sizes and the target distance, the target is less likely to locate at the near-field of both the transmitter and receiver sides. Therefore, in the following, we consider XL-MIMO and XL-phased array sensing when the target locates at the near-filed of the transmitter while at the far-field of the receiver. In this case, the element of the receive steering vector in (\ref{eq3}) can be expressed as$\footnotesize{\ {{ b}_n}(r,\theta ) = {e^{ - j\frac{{2\pi }}{\lambda }(l - n{d_R}\sin \varphi(r,\theta ) )}}.\ }$Based on the expression of$\ \varphi(r,\theta )$ in (\ref{transform}),$\ {{b}_n}(r,\theta )\ $can be written as 
\begin{equation}
	\setlength\abovedisplayskip{2pt}
	\setlength\belowdisplayskip{2pt}
	\footnotesize
	\begin{aligned}
\ {{b}_n}(r,\theta ) = {e^{ - j\frac{{2\pi }}{\lambda }l}}{e^{j\frac{{2\pi }}{\lambda }\frac{{rn{d_R}\sin \theta }}{{\sqrt {{R^2} + {r^2} - 2Rr\cos \theta } }}}} .
	\end{aligned}
\end{equation}
Note that only the phase variation related to $(r,\theta)$ is considered. Therefore, the constant term ${e^{ - j\frac{{2\pi }}{\lambda }l}}$ in ${{b}_n}(r,\theta )$ can be incorporated into the coefficient $\kappa$ in (\ref{eq11}), by defining the new coefficient$\ \hat \kappa  = \kappa {e^{ - j\frac{{2\pi }}{\lambda }l}}.\ $ 
\subsection{XL-MIMO radar}\label{bimimo}
Similar to the analysis in Section \ref{monomimo}, for bistatic XL-MIMO radar mode, the parameters in (\ref{theta}) and (\ref{dis}) can be expressed as
	\begin{equation}
				\setlength\abovedisplayskip{2pt}
		\setlength\belowdisplayskip{1.5pt}
		\footnotesize
				\begin{aligned}
%			\ \begin{array}{l}
				{\| {{{\bf{g}}_\theta }} \|^2} &= Mi + Na + f{c^*} + {f^*}c,\\
				{\| {{{\bf{g}}_r}} \|^2} &= Ms + Np + h{q^*} + {h^*}q,\\
				{\bf{g}}_\theta ^H{{\bf{g}}_r} &= Mk + Ne + f{q^*} + {h^*}c,\\
				{{\bf{g}}^H}{{\bf{g}}_\theta } &= N{c^*} + {f^*}M,
				{\bf{g}}_r^H{\bf{g}} = Nq + Mh,
%			\end{array}\
		\end{aligned}
	\label{inter2}
	\end{equation}
where the intermediate parameters are defined as
\begin{equation}
\setlength\abovedisplayskip{1.5pt}
\setlength\belowdisplayskip{1.5pt}
\footnotesize
\begin{aligned}
i &= {\left\| {\frac{{\partial {\bf{b}}(r,\theta )}}{{\partial \theta }}} \right\|}^2,s = {{\left\| {\frac{{\partial {\bf{b}}(r,\theta )}}{{\partial r}}} \right\|}^2},f = \frac{{\partial {{\bf{b}}^H}(r,\theta )}}{{\partial \theta }}{\bf{b}}(r,\theta ),\\
k &= \frac{{\partial {{\bf{b}}^H}(r,\theta )}}{{\partial \theta }}\frac{{\partial {\bf{b}}(r,\theta )}}{{\partial r}},h = \frac{{\partial {{\bf{b}}^H}(r,\theta )}}{{\partial r}}{\bf{b}}(r,\theta ).
\end{aligned}
\label{bipara}
\end{equation}
%		\begin{align}
%			\ \begin{array}{l}
%				{\| {{{\bf{g}}_\theta }} \|^2} = Mi + Na\\
%				{\| {{{\bf{g}}_r}} \|^2} = Ms + Np\\
%				{\bf{g}}_\theta ^H{{\bf{g}}_r} = Mk + Ne\\
%				{{\bf{g}}^H}{{\bf{g}}_\theta } = N{c^*}\\
%				{\bf{g}}_r^H{\bf{g}} = Nq\\
%				{\sin ^2}\Omega  = 1 - \frac{{{{( {N{c^*}} )}^2}}}{{MN(Mi + Na)}}\\
%				{\sin ^2}\Theta  = 1 - \frac{{{{( {Nq} )}^2}}}{{MN(Ms + Np)}}\\
%				\frac{{\Re \{ {{{\bf{g}}^H}{\bf{Wg}}} \}}}{{{{\| {\bf{g}} \|}^2}}} = \frac{{\Re \{ {{M^2}k + MNe - N{c^*}q} \}}}{M}\\
%				\det Q' = [ {Mi + Na - \frac{N}{M}{{| c |}^2}} ][ {Ms + Np - \frac{N}{M}{{| q |}^2}} ]\\
%				- {[ {Mk + Ne - \frac{N}{M}{c^*}q} ]^2}
%			\end{array}\
%		\end{align}
In order to derive the closed-form expressions for the intermediate parameters in (\ref{bipara}), we define$\ {\Gamma _\theta }(r,\theta )\ $and$\ {\Gamma _r }(r,\theta ),\ $as
	\begin{equation}
		\setlength\abovedisplayskip{1.5pt}
		\setlength\belowdisplayskip{1.5pt}
		\footnotesize
		\begin{aligned}
					\ \begin{array}{*{20}{l}}
				{{\Gamma _\theta }(r,\theta ) \triangleq \frac{{\partial \sin \varphi }}{{\partial \theta }} = \frac{{r\cos \theta ({R^2} + {r^2} - Rr\cos \theta ) - R{r^2}}}{{{{( {{R^2} + {r^2} - 2Rr\cos \theta } )}^{3/2}}}}},\\
				{{\Gamma _r}(r,\theta ) \triangleq \frac{{\partial \sin \varphi }}{{\partial r}} = \frac{{R\sin \theta (R - r\cos \theta )}}{{{{( {{R^2} + {r^2} - 2Rr\cos \theta } )}^{3/2}}}}},
				\label{gamma}
			\end{array}\
		\end{aligned}
	\end{equation}
which are independent of the index of antenna element $n$. Furthermore, by considering the symmetry of antenna array, the parameters in (\ref{bipara}) reduce to
\begin{equation}
\setlength\abovedisplayskip{1pt}
\setlength\belowdisplayskip{1pt}
\footnotesize
\begin{aligned}
i &= \frac{{{\pi ^2}d_R^2\Gamma _\theta ^2}}{{3{\lambda ^2}}}N({N^2} - 1),
s = \frac{{{\pi ^2}d_R^2\Gamma _r^2}}{{3{\lambda ^2}}}N({N^2} - 1),\\
f &= h = 0,
k = \frac{{{\pi ^2}d_R^2{\Gamma _\theta }{\Gamma _r}}}{{3{\lambda ^2}}}N({N^2} - 1).
\end{aligned}
\label{para2}
\end{equation}
%Note that if the receive range is large enough, these parameters can be further simplified and we have $is=k^2$.
\begin{theorem}\label{near-bi}
For bistatic XL-MIMO radar sensing with near-field at the transmitter and far-field at receiver, the closed-form expression of CRBs are
\begin{equation}
\setlength\abovedisplayskip{1.5pt}
\setlength\belowdisplayskip{1.5pt}
\footnotesize
\hspace{-1ex}
\begin{aligned}
	&CR{B_\theta } = \frac{1}{{2\gamma L}}\\
	 &\times \frac{{M( {Ms + Np - \frac{N}{M}{| q |}^2} )}}{{ {( {Mi + Na - \frac{N}{M}{{| c |}^2}} )( {Ms + Np - \frac{N}{M}{{| q |}^2}} ) - {{( {Mk + Ne - \frac{N}{M}{c^*}q} )}^2}} }},
\end{aligned}
\label{bistatic1}
\end{equation}
\begin{equation}
	\setlength\abovedisplayskip{1.5pt}
	\setlength\belowdisplayskip{1.5pt}
	\footnotesize
	\hspace{-1ex}
	\begin{aligned}
	&CR{B_r} = \frac{1}{{2\gamma L}}\\
	 &\times \frac{{M( {Mi + Na - \frac{N}{M}{| c |}^2} )}}{{ {( {Mi + Na - \frac{N}{M}{{| c |}^2}} )( {Ms + Np - \frac{N}{M}{{| q |}^2}} ) - {{( {Mk + Ne - \frac{N}{M}{c^*}q} )}^2}} }},
\end{aligned}
\label{bistatic2}
\end{equation}
where the intermediate parameters $a,c,e,p,q$ are given in closed-form in (\ref{parametera})-(\ref{parameterq}). 
\begin{IEEEproof}
Theorem \ref{near-bi} can be shown by calculating the terms in (\ref{inter2}) as ${\| {{{\bf{g}}_\theta }} \|^2} = Mi + Na,~{\| {{{\bf{g}}_r}} \|^2} = Mj + Np,~{\bf{g}}_\theta ^H{{\bf{g}}_r} = Mk + Ne,~{{\bf{g}}^H}{{\bf{g}}_\theta } = N{c^*},~
{\bf{g}}_r^H{\bf{g}} = Nq.\ $
\end{IEEEproof}
\end{theorem}

The CRB of angle for bistatic MIMO radar based on conventional far-field UPW model is given in \cite{b28}
\begin{equation}
	\setlength\abovedisplayskip{2pt}
	\setlength\belowdisplayskip{2pt}
	\footnotesize
	\begin{aligned}
		{C_\theta } = \frac{1}{{2\gamma L}}\frac{{3{\lambda ^2}}}{{{\pi ^2}N [ {d_R^2( {{N^2} - 1} ) + d_T^2( {{M^2} - 1} )} ]{{\cos }^2\theta}}}.
		\label{upw-bistatic}
	\end{aligned}
\end{equation}
The closed-form CRBs in Theorem \ref{near-bi} are rather involved. To gain some useful insights, we consider the special case when $\theta=0$ in the following. 
\begin{corollary}\label{bistatic}
When $\theta=0$, the CRBs in Theorem \ref{near-bi} reduce to
\begin{equation}
	\setlength\abovedisplayskip{1.5pt}
	\setlength\belowdisplayskip{1.5pt}
	\footnotesize
	\hspace{-1.8ex}
\begin{aligned}
		CR{B_\theta } = \frac{1}{{2\gamma L}}\frac{M}{{\frac{{d_R^2rMN({N^2} - 1)}}{{3{\lambda ^2}(R - r)}} + \frac{{4{\pi ^2}{r^2}N}}{{{\lambda ^2}\varepsilon _T^2}}\left[ {\frac{{{D_T}}}{r} - 2\arctan \left( {\frac{{{D_T}}}{{2r}}} \right)} \right]}},
\end{aligned}
\label{0bistatic1}
\end{equation}
\begin{equation}
	\setlength\abovedisplayskip{1.5pt}
	\setlength\belowdisplayskip{1.5pt}
	\footnotesize
	\begin{aligned}
		&CR{B_r} = \frac{1}{{2\gamma L}}\\
		&\frac{{{\lambda ^2}}}{{4{\pi ^2}N\left[ {\frac{{2r}}{{{D_T}}}\arctan \left( {\frac{{{D_T}}}{{2r}}} \right) - {{\left( {\frac{{2r}}{{{D_T}}}} \right)}^2}\ln^2 \left( {\frac{{{D_T}}}{{2r}} + \sqrt {1 + {{\left( {\frac{{{D_T}}}{{2r}}} \right)}^2}} } \right)} \right]}}.
	\end{aligned}
\label{0bistatic2}
\end{equation}
\begin{IEEEproof}
	Please refer to Appendix F.
\end{IEEEproof}
Corollary \ref{bistatic} shows that the CRB for range is only dependent on $\frac{D_T/2}{r}$, which is determined by the angle between the two lines from the target to the ends of the transmit array, or the transmit angular span as defined in \cite{b2}. Furthermore, it can be shown that the CRB for range in (\ref{0bistatic2}) does not decrease monotonically with the transmit array aperture $D_T$. Instead, it first decreases and then increases as $D_T$ increases, and the minimal point occurs at $D_T\approx12r$, as proved in Appendix F. This indicates that when the number of receive antennas $N$ is fixed, larger transmit aperture $D_T$ does not necessarily lead to better range estimation. This can be explained by the fact that for XL-MIMO radar mode, when the total power of the antenna array is fixed to $P$, the power of each antenna element decreases as the number of antenna increases, as evident from (\ref{eq18}). However, (\ref{derivative1}) and (\ref{derivative2}) show that the magnitude of the partial derivatives decrease for larger antenna index $m$. This means that when $D_T$ is larger enough, the marginal contribution by adding additional antennas fail to compensate the resulting power reduction of each antenna element, which leads to the increase of the CRB for range.

\end{corollary}
\begin{corollary}\label{th8}
When $\theta=0$, for the asymptotic case that $\frac{D_T}{r\cos\theta}\to \infty$, the CRBs of angle and range in (\ref{0bistatic1}) and (\ref{0bistatic2}) approach to the following limits:
\begin{equation}
\setlength\abovedisplayskip{2pt}
\setlength\belowdisplayskip{1pt}
\footnotesize
\hspace{-2.3ex}
\begin{aligned}
 \;\mathop {\lim }\limits_{\frac{{{D_T}}}{{r\cos \theta }} \to \infty } \ CR{B_\theta } = \frac{1}{{2\gamma L}}\frac{{{\lambda ^2}}}{{{\pi ^2}{r^2}N\left[ {\frac{{d_R^2}}{{3{{(R - r)}^2}}}({N^2} - 1) + 4} \right]}},
\end{aligned}
\label{inf-bistatic1}
\end{equation} 
\begin{equation}
	\setlength\abovedisplayskip{1pt}
	\setlength\belowdisplayskip{2pt}
	\footnotesize
	\hspace{-1.5ex}
	\begin{aligned}
	\mathop {\lim }\limits_{\frac{{{D_T}}}{{r\cos \theta }} \to \infty } \;CR{B_r} &= \mathop {\lim }\limits_{\frac{{{D_T}}}{{r\cos \theta }} \to \infty } \frac{1}{{2\gamma L}}\;\frac{{{D_T}^2{\lambda ^2}}}{{4{\pi ^2}{r^2}N\left( {\pi \frac{{{D_T}}}{r} - 4{{\ln }^2}\frac{{{D_T}}}{r}} \right)}}\\
	&\to \infty.
\end{aligned}
\label{inf-bistatic2}
\end{equation} 
\begin{IEEEproof}
Corollary \ref{th8} can be shown by substituting the $g_2$ and $g_4$ in Appendix B into (\ref{0bistatic1}) and (\ref{0bistatic2}), and (\ref{inf-bistatic2}) can be written as
\begin{equation}
	\setlength\abovedisplayskip{1.5pt}
	\setlength\belowdisplayskip{1.5pt}
	\footnotesize
	\begin{aligned}
		CR{B_r}\mathop  \approx \limits^{(a)} \frac{1}{{2\gamma L}}\frac{{{D_T}^2{\lambda ^2}}}{{4{\pi ^2}{r^2}N\pi \frac{{{D_T}}}{r}}}	= \frac{1}{{2\gamma L}}\frac{{{\lambda ^2}}}{{4{\pi ^2}N\pi }}\frac{{{D_T}}}{r}\to\infty.
	\end{aligned}
\end{equation}
with $(a)$ follows from Appendix C.
\end{IEEEproof}
Corollary \ref{th8} shows that when the transmit array aperture $D_T$ goes large, the CRB for angle will approach to a limit that is dependent on the number of receive antennas $N$, rather than decreasing indefinitely with  $D_T$. Besides, the CRB of range depends on both the number of the transmit and receive elements, and will approach to infinity when $D_T$ goes large, which is consistent with Corollary \ref{bistatic}.
\end{corollary} 
\begin{corollary}\label{lemma4}
When $\theta=0$, for the asymptotic case when $\frac{D_T}{r}\ll 1$, the CRBs for angle and range in (\ref{0bistatic1}) and (\ref{0bistatic2}) reduce to
\begin{equation}
\setlength\abovedisplayskip{0pt}
\setlength\belowdisplayskip{1pt}
\footnotesize
\begin{aligned}
CR{B_\theta } &\approx \frac{1}{{2\gamma L}}\frac{{3{\lambda ^2}}}{{{\pi ^2}N\left( {d_R^2{{\left( {\frac{r}{{R - r}}} \right)}^2}({N^2} - 1) + d_T^2{M^2}} \right)}},\\
CR{B_r} &= \frac{1}{{2\gamma L}}\frac{M}{{Np - \frac{N}{M}{{\left| q \right|}^2}}} \to \infty.
\end{aligned}
\label{71}
\end{equation}
\begin{IEEEproof}
Corollary \ref{lemma4} can be shown by substituting the intermediate parameters in Appendix D into (\ref{0bistatic1}) and (\ref{0bistatic2}).
\end{IEEEproof}
Corollary \ref{lemma4} shows that when the array aperture $D_T$ is much smaller than the target range $r$, our newly derived CRB results are consistent with the existing results (\ref{upw-bistatic}) based on the conventional UPW model. Note that the additional factor $(r/(R-r))^2$ in the denominator of (\ref{71}) is due to the fact that we expressed the receiver side angle $\varphi$ in terms of $r$ and $\theta$ in (\ref{transform}), so ${\Gamma _\theta }(r,\theta )$ is different from$\ \frac{{\partial \sin \theta }}{{\partial \theta }}\ $in the far-field UPW model, even when $\theta=\varphi$.
\end{corollary}
\subsection{XL-phased array radar}\label{b1}
For XL-phased array radar mode, the parameters in (\ref{theta}) and (\ref{dis}) can be expressed as
\begin{equation}
\setlength\abovedisplayskip{1.5pt}
\setlength\belowdisplayskip{1.5pt}
\footnotesize
\begin{aligned}
&{\| {{{\bf{g}}_\theta }} \|^2} = i,
{\| {{{\bf{g}}_r}} \|^2} = s,
{\sin ^2}\Omega  = {\sin ^2}\Theta  = 1,\\
&\frac{{\Re \{ {{{\bf{g}}^H}{\bf{Wg}}} \}}}{{{{\| {\bf{g}} \|}^2}}} = \Re \{ k \},
\det \bm Q' = is - {\Re ^2}\{ k \}=0.
\end{aligned}
\end{equation}
Note that since $\det \bm Q' = 0$, it immediately follows from (\ref{theta}) and (\ref{dis}) that both the CRBs for angle and range will be infinity, which implies that in this scenario, neither the angle nor the range can be estimated. This is expected since when far-field UPW model is applied at the receiver side and phased array beamforming is applied at the transmitter side, the considered bistatic sensing problem is equivalent to the far-field target localization problem \cite{b80}. In this case, only the angle $\varphi$ of the target with respect to the receive array can be estimated, and it is impossible to obtain the transmit angle $\theta$ and range $r$ based on $\varphi$ alone. Therefore, both the CRBs for angle and range is infinity.
%\subsection{Near-field Receiver}
%In this subsection, we consider bistatic XL-MIMO and XL-phased array sensing when the target locates at the far-field of the transmitter while at the near-field of the receiver.
%\subsubsection{XL-MIMO radar}
%In this case, the output of the matched filter is similar to that for the monostatic case in Section \ref{monomimo}, and the only difference is that the parameters to be estimated is $l$ and $\varphi$, instead of $r$ and $\theta$. Therefore, the CRBs of angle and range under this scenario can be directly obtained by replacing $r$ and $\theta$ in (\ref{bistatic1})-(\ref{bistatic2}) with $l$ and $\varphi$, and the CRBs of $\theta$ and $r$ can be obtained based on (\ref{transform}).
%\subsubsection{XL-phased array radar}
%Similar to the analysis in section \ref{b1}, the bistatic phased array radar sensing is equivalent to a target localization problem. Therefore, the closed-form expressions of CRBs for angle $\varphi$ and range $l$ are similar to the monostatic case in (\ref{crb-ph1}) and (\ref{crb-ph2}) by changing the parameters to be estimated from $\theta$ and $r$ to $\varphi$ and $l$. The detailed derivations are omitted for brevity. 

To sum up, the CRBs for different classes of sensing discussed above are given in Table \uppercase\expandafter{\romannumeral1}.		
\begin{table*}
	\label{table1}
	\centering
	\renewcommand\arraystretch{2}
	\hspace{-6ex}
	\caption{CRBs for different classes of Sensing under various scenarios}
	\begin{tabular}{|c|c|c|c|c|c|c|} 
		
		\hline
		&                             &                           & \multicolumn{2}{c|}{XL-MIMO radar mode}                               & \multicolumn{2}{c|}{XL-phased array radar mode}                  \\ 
		\hline
		&                             &                           & $CRB_\theta$ & $CRB_r$                             & $CRB_\theta$ & $CRB_r$                        \\	
		\hline
		\multirow{5}{*}{Monostatic } & \multirow{3}{*}{\makecell[c]{Near-Field\\ USW model}}
		& General result                     & {(\ref{crb-mimo1})}     & {(\ref{crb-mimo2})}  &(\ref{crb-ph1}) &(\ref{crb-ph2})  \\ 
		\cline{3-7}
		&                              & \makecell[c]{$D_T\gg r$}  
		& {(\ref{extt})}     & {(\ref{extr})}  &(\ref{extt2}) &(\ref{extr2})  \\
		\cline{3-7}
		&                             & \makecell[c]{$\frac{D_T}{r\cos\theta} \to \infty$} 
		&$\frac{1}{{2\gamma L}}\frac{{{\lambda ^2}{d_T}{{\sin }^2}\theta }}{{8{\pi ^3}{r^3}\cos \theta }}$&$\frac{1}{{2\gamma L}}\frac{{{\lambda ^2}{d_T}\cos \theta }}{{8{\pi ^3}r}}$ & $\frac{1}{{2\gamma L}}\frac{{{\lambda ^2}{d_T}{{\sin }^2}\theta }}{{4M{\pi ^3}{r^3}\cos \theta }}$& $\frac{1}{{2\gamma L}}\frac{{{\lambda ^2}{d_T}\cos \theta }}{{4M{\pi ^3}r}}$  \\ 
		\cline{2-7}
		& \multicolumn{2}{c|}{\makecell[c]{Near-Field USW model with\\ second-order Taylor approximation}             }              & 
		{$\frac{1}{{2\gamma L}}\frac{{3{\lambda ^2}}}{{2{\pi ^2}d_T^2M ( {{M^2} - 1} ){{\cos }^2}\theta}}$} & {(\ref{crb-taylor-mimo2})} & {$\frac{1}{{2\gamma L}}\frac{{3{\lambda ^2}}}{{{\pi ^2}d_T^2M^2( {{M^2} - 1} ){{\cos }^2}\theta }}$}  & {(\ref{crb-taylor-ph2})}\\ 
		\cline{2-7}
		& \multicolumn{2}{c|}{\makecell[c]{Far-Field UPW model\\ $D_T \ll r$\cite{b28}}             }              & 
		{$\frac{1}{{2L{\gamma }}}\frac{{3{\lambda ^2}}}{{2{\pi ^2}d_T^2M( {{M^2} - 1} ){{\cos }^2}\theta }}$} & {$\infty$} & {$\frac{1}{{2\gamma L}}\frac{{3{\lambda ^2}}}{{{\pi ^2}d_T^2{M^2} ({M^2} - 1){{\cos }^2}\theta}}\ $}  & {$\infty$}\\ 
		\hline
		\multirow{3}{*}{\makecell[c]{Bistatic  }}   & \multirow{2}{*}{\makecell[c]{Near-Field \\transmitter\\ USW model}} & General result                   & (\ref{bistatic1})&(\ref{bistatic2})    & \multicolumn{2}{c|}{$\infty$}                                \\ 
		\cline{3-7}
		&                             & \makecell[c]{$\frac{D_T}{r\cos\theta} \to \infty$} &(\ref{inf-bistatic1})&(\ref{inf-bistatic2}) & \multicolumn{2}{c|}{$\infty$}                                \\ 
		\cline{2-7}
		& \multicolumn{2}{c|}{Far-Field UPW model\cite{b66}}                          & (\ref{upw-bistatic})  &{$\infty$}  & {\cite{b66}}    &{$\infty$}                            \\ 
		\hline
	\end{tabular}
\end{table*}
\section{Numerical results}{\label{simul}}
Numerical results are provided in this section to validate our derived near-field CRB results. Unless otherwise stated, the carrier frequency is $f=2.37$GHz, $d_T=d_R=0.0628$m, and the SNR is$\ \gamma=\frac{{{P{| \kappa  |}^2}}}{{{\sigma ^2}}}=0\ $dB.
\subsection{Monostatic Sensing}
 \begin{figure}[htbp]
	\setlength{\abovecaptionskip}{-0.2cm}
\setlength{\belowcaptionskip}{-0.5cm}
	\centerline{\includegraphics[width=0.45\textwidth]{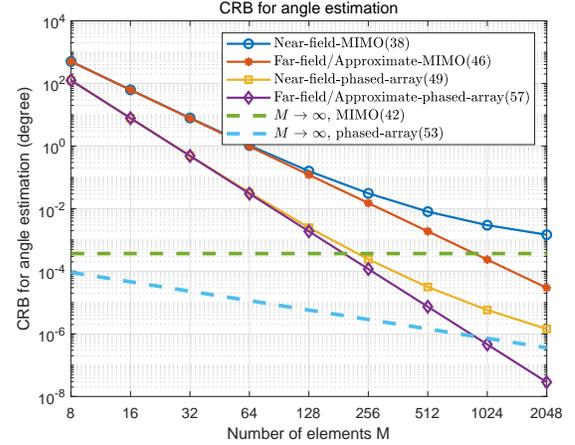}}
	\caption{CRB of angle for monostatic sensing.}
	\label{angle-crb}
\end{figure}
\begin{figure}[htbp]
	\setlength{\abovecaptionskip}{-0.2cm}
\setlength{\belowcaptionskip}{-0.5cm}
	\centerline{\includegraphics[width=0.45\textwidth]{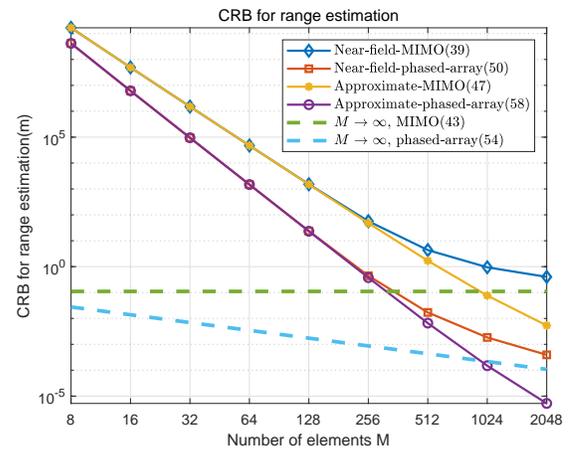}}
	\caption{CRB of range for monostatic sensing. For conventional UPW model, CRB of range is $\infty$.}
	\label{range-crb}
\end{figure}

Fig.\ref{angle-crb} and Fig.\ref{range-crb} plot the CRBs of angle $\theta$ and range $r$ versus the number of antennas $M$ for various models, as given in (\ref{crb-mimo1})-(\ref{crb-mimo2}), (\ref{crb-ph1})-(\ref{crb-ph2}), (\ref{crb-mimo-inf1}) and (\ref{inf-ph1}), respectively. In this case, the target range is set to be $r=10$m, while the angle is $\theta=\pi/6$. The legend ``Near-field-MIMO" and ``Near-field-phased array" denote the general expression of CRBs in (\ref{crb-mimo1}), (\ref{crb-mimo2}), (\ref{crb-ph1}) and (\ref{crb-ph2}), while ``Approximate-MIMO" and ``Approximate-phased array" denote the CRBs based on the second-order Taylor approximation in (\ref{crb-taylor-mimo1}), (\ref{crb-taylor-mimo2}), (\ref{crb-taylor-ph1}) and (\ref{crb-taylor-ph2}), respectively. Besides, ``Far-field-MIMO" and ``Far-field-phased array" denote CRBs of the conventional UPW model, respectively. The dotted lines are the asymptotic limits when $\frac{D_T}{r\cos\theta}\to \infty$.
Fig.\ref{angle-crb} shows that the CRBs for angle of phased array radar mode is smaller than that of MIMO radar mode, since the former usually benefits from an additional transmit beamforming gain. Besides, the CRBs of angle for both XL-MIMO radar and XL-phased array radar decrease with the increase of antenna size, but with diminishing return. Furthermore, for relatively small $M$ values, the CRBs of angle for XL-MIMO and XL-phased array radar are consistent with the far-field CRBs. However, with the increasing of antenna number, the two models lead to dramatical different results. This indicates that using inappropriate far-field model to analyse near-field sensing with extremely large-scale arrays may cause severe errors.
Furthermore, it is observed from Fig. \ref{range-crb} that the range CRBs also decrease with the number of antenna elements. Besides, for moderate antenna number, say $M<64$, the CRBs of range are quite large. It is observed that the second-order Taylor approximation as (\ref{crb-taylor-mimo1})-(\ref{crb-taylor-mimo2}) and (\ref{crb-taylor-ph1})-(\ref{crb-taylor-ph2}) is accurate for moderate $M$, but would lead to large errors when $M$ is large. 

\begin{figure}[!htbp]
	\setlength{\abovecaptionskip}{-0.2cm}
	\setlength{\belowcaptionskip}{-0.5cm}
	\centerline{\includegraphics[width=0.45\textwidth]{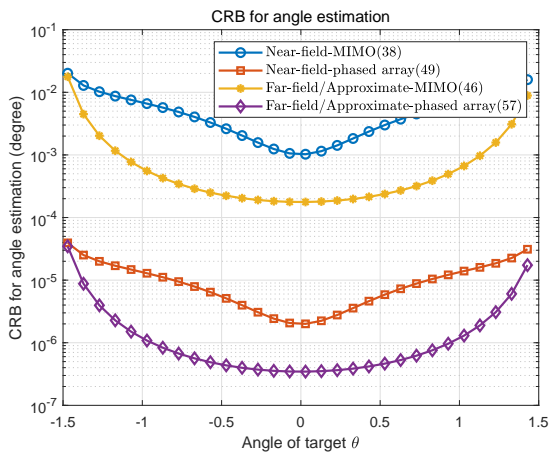}}
	\caption{CRB of angle for monostatic sensing.}
	\label{mono_theta_t}
\end{figure}
\begin{figure}[!htbp]
	\setlength{\abovecaptionskip}{-0.2cm}
	\setlength{\belowcaptionskip}{-0.5cm}
	\centerline{\includegraphics[width=0.45\textwidth]{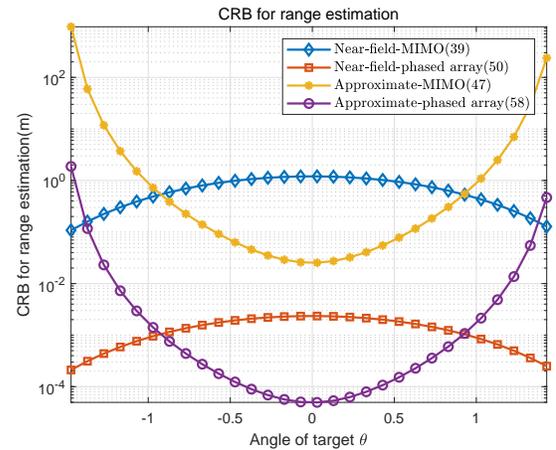}}
	\caption{CRB of range for monostatic sensing.}
	\label{mono_theta_r}
\end{figure}
Fig.\ref{mono_theta_t} and Fig.\ref{mono_theta_r} plot the CRBs of angle and range versus the target angle $\theta$. The number of antenna elements is set to be $M=1024$, while the target range is $r=10$m. Fig.\ref{mono_theta_t} shows that the CRB of angle increases with$\ | \theta  |,\ $and the largest CRB occurs at the boresight of the array. Besides, for the considered setup, using second-order Taylor approximation or far-field UPW model may cause $20$dB error on average, which is nonnegligible in practice. 
Furthermore, Fig.\ref{mono_theta_r} shows that CRBs based on the second-order Taylor approximation have the opposite trend compared with the CRBs based on the exact distances, which indicates that using second-order Taylor approximation would be inaccurate for near-field CRB for range. 

\begin{figure}[!htbp]
	\setlength{\abovecaptionskip}{-0.2cm}
	\setlength{\belowcaptionskip}{-0.7cm}
	\centerline{\includegraphics[width=0.45\textwidth]{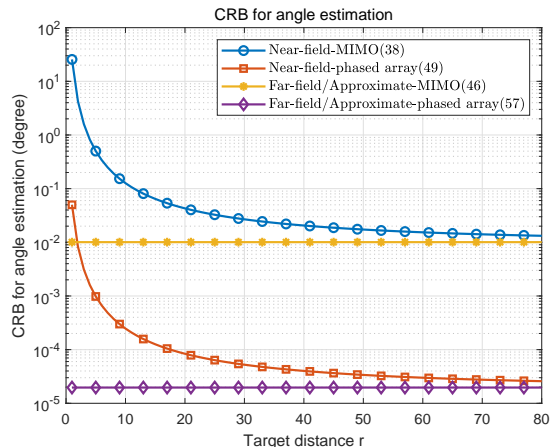}}
	\caption{CRB of angle for monostatic sensing.}
	\label{mono_range_t}
\end{figure}
\begin{figure}[!htbp]
	\setlength{\abovecaptionskip}{-0.2cm}
	\setlength{\belowcaptionskip}{-0.9cm}
	\centerline{\includegraphics[width=0.45\textwidth]{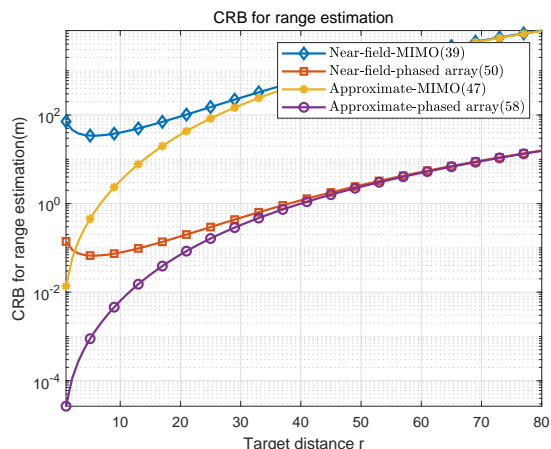}}
	\caption{CRB of range for monostatic sensing.}
	\label{mono_range_r}
\end{figure}
Fig.\ref{mono_range_t} and Fig.\ref{mono_range_r} plot the CRBs of angle and range versus the target range $r$. The number of antenna elements is set to be $M=1024$, while target angle is $\theta=\pi/6$. 
Fig.\ref{mono_range_t} shows that the CRB of angle decreases with the increasing of target range. Besides, for large target range $r$, the newly developed near-field CRB matches with that from the conventional UPW models. However, they deviate significantly for relatively small $r$, where far-field UPW assumption no longer holds. Fig.\ref{mono_range_r} shows that the developed near-field CRB for range merges with the conventional result when the range $r$ is large. Besides, as $r$ increases, the derived CRB of range based on the exact distance first decreases and then increases, while for second-order Taylor approximation model, the CRB increases monotonically. This shows that the second-order Taylor approximation model is less accurate for smaller target range $r$.
\subsection{Bistatic Sensing}
In this subsection, bistatic near-field sensing results are presented. corresponding to section \ref{bimimo}, near-field USW model is considered only at the transmitter side and the number of the receive antenna elements is set to be $N=8$. The target angle is $\theta=0$, and the target range is $r=18$m. The transmit and receive array distance is $R=35$m. 
The classic Capon algorithm is used to actually estimate the parameters $\theta$ and $r$, and the performance is evaluated in terms of the root mean square error (RMSE) in the following way\cite{b53}
\begin{equation}
	\setlength\abovedisplayskip{1pt}
	\setlength\belowdisplayskip{1.5pt}
	\footnotesize
	\begin{aligned}
\ RMSE_i = \sqrt {\frac{1}{K}\sum\limits_{k = 1}^K {{{( {{\bm {\theta}_i } - {\bm{\hat \theta }_i}} )}^2}} },i=1,2,\
	\end{aligned}
\end{equation}
where$\ {\bm{\theta }_i} = [\theta ,r],\ $and $\bm{\hat \theta }_i$ denotes the estimate value of $\theta$ and $r$. $K$ is the total number of experiment, and we set $K=500$. 
\begin{figure}[htbp]
	\setlength{\abovecaptionskip}{-0.1cm}
	\setlength{\belowcaptionskip}{-0.3cm}
	\centerline{\includegraphics[width=0.45\textwidth]{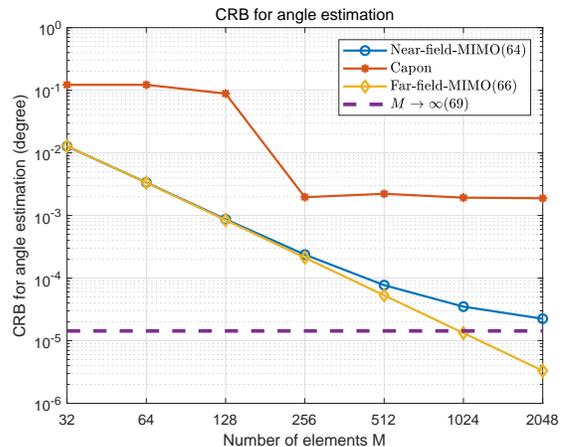}}
	\caption{CRB of angle for bistatic sensing.}
	\label{bi_M_t}
\end{figure}
Fig.\ref{bi_M_t} and Fig.\ref{bi_M_r} plot the CRBs of angle and range versus the transmit antenna number $M$. It is observed that our derived CRBs are indeed the lower bounds for the RMSE of the Capon algorithms. Besides, the developed near-field CRB perfectly match with the far-field model when $M$ is relatively small.
% Fig.\ref{bi_M_r} also shows the turning point where the CRB of range for bistatic XL-MIMO reaches the minimum, which verifies the result in Corollary \ref{bistatic}.
\begin{figure}[htbp]
	\setlength{\abovecaptionskip}{-0.1cm}
	\setlength{\belowcaptionskip}{-0.3cm}
	\centerline{\includegraphics[width=0.45\textwidth]{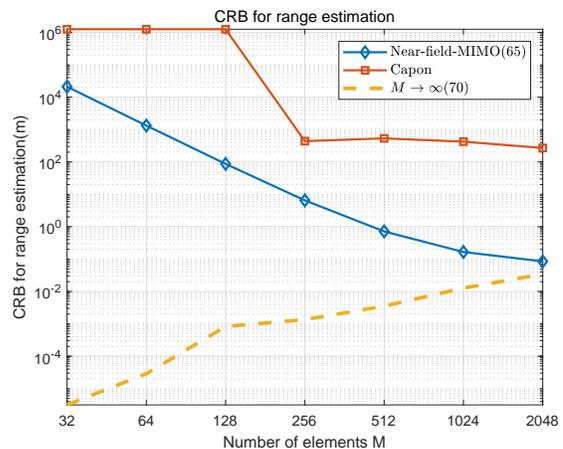}}
	\caption{CRB of range for bistatic sensing.}
	\label{bi_M_r}
\end{figure}
\section{Conclusion}\label{sec:Conclusion}
This paper studied near-field radio sensing with extremely large-scale antenna arrays, where the USW model was considered. We considered two radar modes, namely XL-MIMO radar and XL-phased array radar modes. For the monostatic near-field sensing, the closed-form expressions of the CRBs for angle and range estimation were derived. Several asymptotic cases were also considered. It was revealed that different from the conventional far-field sensing with UPW model, the CRB for near-field angle estimation no longer decreases indefinitely as the antenna size increases. Moreover, the CRB for range estimation was shown to be finite in the near-field case, which shows the capability for range discrimination with XL-MIMO sensing. We then considered the general bistatic case and compared our derived CRBs with the classic Capon algorithm. Numerical results validated our derived CRBs. 
\section*{Appendix A\\Proof of Proposition \ref{prop1}}
As shown in (\ref{parametera})-(\ref{parameterq}), the different parameters share some common summations. Therefore, similar to \cite{b2}, in order to obtain closed-form expressions, we first define the function$\ {f_1}(x) \buildrel \Delta \over = \frac{{{x^2}}}{{{x^2} - 2(\sin \theta) x + 1}},\  $where$\ x \in [ { - \frac{{M{\varepsilon _T}}}{2},\frac{{M{\varepsilon _T}}}{2}} ].\ $Since$\ {\varepsilon _T} \ll 1,\ $we have$\ f_1(x) \approx f_1(m{\varepsilon _T}),~\forall x \in [ {( {m - \frac{1}{2}} ){\varepsilon _T},( {m + \frac{1}{2}} ){\varepsilon _T}} ],~m =  - \frac{{M - 1}}{2},...,\frac{{M - 1}}{2}.\ $So we have
\begin{equation}
	\setlength\abovedisplayskip{0pt}	\setlength\belowdisplayskip{2pt}
	\footnotesize
\begin{aligned}
		\ \sum\nolimits_{m =  - \frac{{M - 1}}{2}}^{\frac{{M - 1}}{2}} {f_1(m{\varepsilon _T}){\varepsilon _T}} \approx \int_{ - \frac{{M{\varepsilon _T}}}{2}}^{\frac{{M{\varepsilon _T}}}{2}} {f_1(x)dx}.\
\end{aligned}
	\label{eq4}
\end{equation}
Therefore, the summation (\ref{psuma}) and (\ref{psump}) can be expressed as
\begin{equation}
				\setlength\abovedisplayskip{1.5pt}
	\setlength\belowdisplayskip{1.5pt}
	\footnotesize
%	\ \begin{aligned}{*{20}{l}}
	\begin{aligned}
	&\sum\nolimits_{m =  - \frac{{M - 1}}{2}}^{\frac{{M - 1}}{2}} {\frac{{{{( {m{\varepsilon _T}} )}^2}}}{{1 - 2m{\varepsilon _T}\sin \theta  + {{(m{\varepsilon _T})}^2}}}} \\
	&\approx \frac{1}{{{\varepsilon _T}}}\int_{ - \frac{{M{\varepsilon _T}}}{2}}^{\frac{{M{\varepsilon _T}}}{2}} {\frac{{{x^2}}}{{{x^2} - 2(\sin \theta) x + 1}}} dx\\
%		\end{aligned}}\\
&= M + \frac{{\sin \theta }}{{{\varepsilon _T}}}\int_{ - \frac{{M{\varepsilon _T}}}{2}}^{\frac{{M{\varepsilon _T}}}{2}} {\frac{1}{{{x^2} - 2(\sin \theta) x + 1}}} d( {{x^2} - 2(\sin \theta) x + 1} )\\
		&-\frac{\cos2\theta}{{{\varepsilon _T}}}\int_{ - \frac{{M{\varepsilon _T}}}{2}}^{\frac{{M{\varepsilon _T}}}{2}} {\frac{1}{{{x^2} - 2(\sin \theta) x + 1}}} dx\\
	&\mathop  = \limits^{(a)} M + \frac{{\sin \theta }}{{{\varepsilon _T}}}\ln \left| {\frac{{\frac{{D_T^2}}{{4{r^2}}} - \sin \theta\frac{{{D_T} }}{r} + 1}}{{\frac{{D_T^2}}{{4{r^2}}} + \sin \theta\frac{{{D_T} }}{r} + 1}}} \right| - \frac{{\cos 2\theta }}{{{\varepsilon _T}\cos \theta }}\Delta _{{\rm{span}}}^{\rm{t}}\left(\frac{{{D_T}}}{r}\right),
	\end{aligned}\
	\label{sum1}
\end{equation}
where$\ \Delta _{{\rm{span}}}^t(\frac{{{D_T}}}{r}) = \arctan ( {\frac{{{D_T}}}{{{\rm{2}}r\cos \theta }} - \tan \theta } ) + \arctan ( {\frac{{{D_T}}}{{{\rm{2}}r\cos \theta }} + \tan \theta } )\ $is the transmit angular span\cite{b2}. $(a)$ in (\ref{sum1}) follows from the integral formula 2.103 in \cite{b76}, i.e.,$\ \int {\frac{{(Mx + N)dx}}{{A + 2Bx + C{x^2}}} = \frac{M}{{2C}}\ln \left| {A + 2Bx + C{x^2}} \right|}  + \frac{{NC - MB}}{{C\sqrt {AC - {B^2}} }}\arctan \frac{{Cx + B}}{{\sqrt {AC - {B^2}} }}\ $for$\ AC > {B^2},\ $and the fact that $\sin^2\theta\le1$ in (\ref{sum1}). Therefore, the closed-form expression of the parameter $a$ and $p$ can be derived by substituting (\ref{sum1}) in (\ref{parametera}) and (\ref{parameterp}).

Similarly, we define the functions$\ {f_2}(x) \buildrel \Delta \over = \frac{x}{{\sqrt {{x^2} - 2(\sin \theta) x + 1} }}\ $and$\ {f_3}(x){\rm{ }} \buildrel \Delta \over = \frac{x}{{{x^2} - 2(\sin \theta) x + 1}},\ $and the summation in (\ref{psumc}) and (\ref{psumq}) can be expressed as
\begin{equation}
\setlength\abovedisplayskip{1pt}
\setlength\belowdisplayskip{1pt}
\footnotesize
\begin{aligned}
&\sum\nolimits_{m =  - \frac{{M - 1}}{2}}^{\frac{{M - 1}}{2}} {\frac{{m{\varepsilon _T}}}{{\sqrt {1 - 2m{\varepsilon _T}\sin \theta  + {{(m{\varepsilon _T})}^2}} }}} \\
&\approx \frac{1}{{{\varepsilon _T}}}\int_{ - \frac{{M{\varepsilon _T}}}{2}}^{\frac{{M{\varepsilon _T}}}{2}} {\frac{x}{{\sqrt {{x^2} - 2\sin (\theta )x + 1} }}} dx\\
&= \frac{1}{{{\varepsilon _T}}}\left[ {\sqrt {\frac{{D_T^2}}{{4{r^2}}} - \sin \theta\frac{{{D_T} }}{r} + 1} }  - {\sqrt {\frac{{D_T^2}}{{4{r^2}}} + \sin \theta\frac{{{D_T} }}{r} + 1} } \right.\\
&\left.+  \psi \left(\frac{D_T}{r}\right)\sin \theta \right],
	\label{sum2}
\end{aligned}
\end{equation}
where$~\psi(\frac{{{D_T}}}{r})\triangleq \ln \left(\frac{{{p_2} + \sqrt {1 + p_2^2} }}{{{p_1} + \sqrt {1 + p_1^2} }}\right)$ with$\ {p_1} = \frac{{ - \frac{{{D_T}}}{{2r}} - \sin \theta }}{{\cos \theta }},~{p_2} = \frac{{\frac{{{D_T}}}{{2r}} - \sin \theta }}{{\cos \theta }},\ $and the summation in (\ref{psume})
\begin{equation}
	\setlength\abovedisplayskip{2pt}
	\setlength\belowdisplayskip{1pt}
	\footnotesize
	\begin{aligned}
	&\sum\nolimits_{m =  - \frac{{M - 1}}{2}}^{\frac{{M - 1}}{2}} {\frac{{m{\varepsilon _T}}}{{1 - 2m{\varepsilon _T}\sin \theta  + {{(m{\varepsilon _T})}^2}}}} \\
	&\approx \frac{1}{{{\varepsilon _T}}}\int_{ - \frac{{M{\varepsilon _T}}}{2}}^{\frac{{M{\varepsilon _T}}}{2}} {\frac{x}{{{x^2} - 2\sin (\theta )x + 1}}} dx\\
	&= \frac{1}{{2{\varepsilon _T}}}\ln \left| {\frac{{\frac{{D_T^2}}{{4{r^2}}} - \sin \theta\frac{{{D_T} }}{r} + 1}}{{\frac{{D_T^2}}{{4{r^2}}} + \sin \theta\frac{{{D_T} }}{r} + 1}}} \right| + \frac{{\sin \theta }}{{{\varepsilon _T}\cos \theta }}\Delta _{{\rm{span}}}^{\rm{t}}\left(\frac{{{D_T}}}{r}\right).
		\label{sum3}
	\end{aligned}
\end{equation}
By substituting (\ref{sum1}) and (\ref{sum3}) into (\ref{parametere}) and (\ref{parameterp}), the closed-form expressions of parameters $e$ and $p$ can be obtained accordingly. Furthermore, by substituting (\ref{sum2}) into (\ref{parameterc}) and (\ref{parameterq}), $c$ and $q$ in Proposition \ref{prop1} can be obtained accordingly. 
\section*{Appendix B\\Proof of Corollary \ref{t1}}
By noting that the parameters in (\ref{parametera})-(\ref{parameterq}) involve some common terms, we first define the following four functions:
\begin{equation}
\setlength\abovedisplayskip{1pt}
\setlength\belowdisplayskip{1pt}
\scriptsize
\begin{aligned}
		{g_1}\left( {\frac{{{D_T}}}{r}} \right) &\triangleq  \sqrt {\frac{{{D_T}^2}}{{4{r^2}}} - \frac{{{D_T}}}{r}\sin \theta  + 1}  - \sqrt {\frac{{{D_T}^2}}{{4{r^2}}} + \frac{{{D_T}}}{r}\sin \theta  + 1} ,\\
		{g_2}\left( {\frac{{{D_T}}}{r}} \right) &\triangleq  \Delta _{{\rm{span}}}^{\rm{t}}\left(\frac{{{D_T}}}{r}\right),
		{g_3}\left( {\frac{{{D_T}}}{r}} \right) \triangleq  \ln \left | {\frac{{\frac{{{D_T}^2}}{{4{r^2}}} - \frac{{{D_T}}}{r}\sin \theta  + 1}}{{\frac{{{D_T}^2}}{{4{r^2}}} + \frac{{{D_T}}}{r}\sin \theta  + 1}}} \right|,\\
		{g_4}\left( {\frac{{{D_T}}}{r}} \right) &\triangleq  \ln \left(\frac{{{p_2} + \sqrt {1 + p_2^2} }}{{{p_1} + \sqrt {1 + p_1^2} }}\right),
\end{aligned}
\end{equation}
where $\ {p_1} = \frac{{ - \frac{{{D_T}}}{{2r}} - \sin \theta }}{{\cos \theta }},\ $and$\ {p_2} = \frac{{\frac{{{D_T}}}{{2r}} - \sin \theta }}{{\cos \theta }},\ $as given in Proposition \ref{prop1}. When $\frac{{{D_T}}}{r}\gg1,\ $as in Corollary \ref{t1}, we have
\begin{equation}
	\setlength\abovedisplayskip{1pt}
	\setlength\belowdisplayskip{1pt}
\scriptsize
%	\hspace{-2.3ex}
	\begin{aligned}
		{{g_1}\left( {\frac{{{D_T}}}{r}} \right)} &= \frac{{{D_T}}}{{2r}} {\left( {\sqrt {1 + \frac{{4{r^2}}}{{{D_T}^2}} -  \frac{4r\sin \theta }{{{D_T}}}} - \sqrt {1 + \frac{{4{r^2}}}{{{D_T}^2}} +  \frac{4r\sin \theta}{{{D_T}}}} } \right) }\\
		&\mathop  \approx \limits^{(a)} \frac{{{D_T}}}{{2r}} \left(-  \frac{4r}{{{D_T}}}\sin \theta\right)
		=  - 2\sin \theta ,
		\label{g1}
	\end{aligned}
\end{equation}
where $(a)$ in (\ref{g1}) follows from the first-order Taylor approximation with $\frac{r}{D_T}\ll1$. Besides, it was shown in \cite{b2} that$\ {g_2}\left( {\frac{{{D_T}}}{r}} \right) = \Delta _{{\rm{span}}}^{\rm{t}}(\frac{{{D_T}}}{r}) \approx \pi .\ $Similarly,$\ {g_3}\left( {\frac{{{D_T}}}{r}} \right)\ $can be obtained as
\begin{equation}
	\setlength\abovedisplayskip{1pt}
	\setlength\belowdisplayskip{1pt}
\footnotesize
	\begin{aligned}
		{g_3}\left( {\frac{{{D_T}}}{r}} \right) &= \ln \left| {1 - \frac{{2\frac{{{D_T}}}{r}\sin \theta }}{{\frac{{{D_T}^2}}{{4{r^2}}} + \frac{{{D_T}}}{r}\sin \theta  + 1}}} \right|
	\approx  0.
	\end{aligned}\
\end{equation}
Furthermore, the last term$\ {g_4}\left( {\frac{{{D_T}}}{r}} \right)\ $can be expressed as
\begin{equation}
	\setlength\abovedisplayskip{1pt}
	\setlength\belowdisplayskip{1pt}
\footnotesize
	\begin{aligned}
		{{g_4}\left( {\frac{{{D_T}}}{r}} \right) }&\mathop  \approx \limits^{(d)} \ln \left( {\frac{{\frac{{{D_T}}}{{2r\cos \theta }} + \sqrt {1 + {{( {\frac{{{D_T}}}{{2r\cos \theta }}} )}^2}} }}{{ - \frac{{{D_T}}}{{2r\cos \theta }} + \sqrt {1 + {{( {\frac{{{D_T}}}{{2r\cos \theta }}} )}^2}} }}} \right)\approx 2\ln \left( {\frac{{{D_T}}}{{r\cos \theta }}} \right),
	\end{aligned}
\end{equation}
where $(d)$ follows from$\ \frac{{\frac{{{D_T}}}{{2r}} - \sin \theta }}{{\cos \theta }} \approx \frac{{{D_T}}}{{2r\cos \theta }}\ $when $D_T/r\gg1.$

Note that the parameter $a$ in (\ref{parametera}) can be written based on $g_2$ and $g_3$ defined above. Therefore, when $D_T/r\gg1,$ by substituting $g_2$ and $g_3$ into (\ref{parametera}), we have
\begin{equation}
\setlength\abovedisplayskip{1.5pt}
\setlength\belowdisplayskip{1.5pt}
\footnotesize
\begin{aligned}
		a &= \frac{{4{\pi ^2}{r^2}{{\cos }^2}\theta }}{{{\lambda ^2}{\varepsilon _T}}}\left( {\frac{{{D_T}}}{r} + {g_3}\sin \theta  - {g_2}\frac{{\cos 2\theta }}{{\cos \theta }}} \right)\\
		&= \frac{{4{\pi ^2}{r^2}{{\cos }^2}\theta }}{{{\lambda ^2}{\varepsilon _T}}}\left( {\frac{{{D_T}}}{r} - \frac{{\cos 2\theta  }}{{\cos \theta }}\pi} \right).
\end{aligned}
\end{equation}
Similarly, other parameters $e,c,p,q$ can be written based on $g_1,g_2,g_3$ and $g_4$ and when $D_T/r\gg1$, we have
\begin{equation}
	\setlength\abovedisplayskip{1.5pt}
	\setlength\belowdisplayskip{1.5pt}
\footnotesize
\begin{aligned}
	e &= \frac{{2{\pi ^2}r\sin 2\theta }}{{{\lambda ^2}{\varepsilon _T}}}\left( {\frac{{{D_T}}}{r} - 2\pi\cos \theta  } \right),\\
	p &= \frac{{4{\pi ^2}}}{{{\lambda ^2}{\varepsilon _T}}}\left( {\frac{{{D_T}}}{r}{{\sin }^2}\theta  + \pi\cos \theta \cos 2\theta  } \right),\\
c &=  - j\frac{{2\pi r\sin 2\theta }}{{\lambda {\varepsilon _T}}}\left[ {\ln \left( {\frac{{{D_T}}}{{r\cos \theta }}} \right) - 1} \right],\\
	q &= j\frac{{4\pi }}{{\lambda {\varepsilon _T}}}\left[ {{{\cos }^2}\theta \ln \left( {\frac{{{D_T}}}{{r\cos \theta }}} \right) + {{\sin }^2}\theta } \right].
\end{aligned}
\end{equation}
%Therefore, the major terms in (\ref{crb-mimo1})-(\ref{crb-mimo2}) can be expressed as
%\begin{equation}
%	\setlength\abovedisplayskip{1.5pt}
%	\setlength\belowdisplayskip{1.5pt}
%\footnotesize
%	\begin{aligned}
%			{Mp - {{| q |}^2}} &\approx \frac{{4{\pi ^2}}}{{{\lambda ^2}{\varepsilon _T}^2}}\left[ { {{\left( {\frac{{{D_T}}}{r}} \right)}^2{{\sin }^2}\theta} +   \frac{{\pi{D_T}}}{r}}\cos \theta \cos 2\theta \right.\\
%			&\left. { - 4{\left( {{{\cos }^2}\theta \ln \frac{{{D_T}}}{{r\cos \theta }} + {{\sin }^2}\theta } \right)^2}} \right],\\
%			{Ma - {{| c |}^2}} &\approx \frac{{4{\pi ^2}{r^2}{{\cos }^2}\theta }}{{{\lambda ^2}{\varepsilon _T}^2}}\left[ {{{\left( {\frac{{{D_T}}}{r}} \right)}^2} -  \frac{{\pi{D_T}}}{r}}\frac{{\cos 2\theta }}{{\cos \theta }} \right.\\
%			&\left.{ - 4 {{\left( {\ln \frac{{{D_T}}}{{r\cos \theta }} - 1} \right)}^2}}{{\sin }^2}\theta \right].
%	\end{aligned}
%\end{equation}

By substituting these terms into (\ref{crb-mimo1}) and (\ref{crb-mimo2}), the expressions in (\ref{extt}) and (\ref{extr}) can be obtained accordingly. 

\section*{Appendix C\\Proof of Corollary \ref{t2}}
When $\frac{D_T}{r\cos\theta} \to \infty$, the constant terms in (\ref{extt}) and (\ref{extr}) can be neglected. Therefore, we have the simplified results in (\ref{crb-mimo-inf1}) and (\ref{crb-mimo-inf2}) by some simple algebraic calculation:
\begin{equation}
			\setlength\abovedisplayskip{1.5pt}
	\setlength\belowdisplayskip{1.5pt}
	\footnotesize
	\begin{aligned}
	CR{B_\theta } &\approx \frac{1}{{2\gamma L}}\\
	&\frac{{{\lambda ^2}\left[\frac{{{{\sin }^2}2\theta }}{4}{{(\frac{{{D_T}}}{{r\cos \theta }})}^2} +  \frac{{\pi {D_T}}}{{r\cos \theta }}{{\cos }^2}\theta \cos 2\theta - 4{{({{\cos }^2}\theta \ln \frac{{{D_T}}}{{r\cos \theta }})}^2}\right]}}{{8{\pi ^2}M{r^2}{{\cos }^2}\theta (\frac{{\pi {D_T}}}{r{\cos \theta}} - 4{{\ln }^2}\frac{{{D_T}}}{{r\cos \theta }})}}\\
	&\mathop  \approx \limits^{(g)} \frac{1}{{2\gamma }}\frac{{{\lambda ^2} {{(\frac{{{D_T}}}{r})}^2}{{\sin }^2}\theta}}{{8{\pi ^2}{r^2}M{{\cos }^2}\theta \frac{{\pi {D_T}}}{r{\cos \theta}}}} \approx \frac{1}{{2\gamma }}\frac{{{\lambda ^2}{d_T}{{\sin }^2}\theta }}{{8{\pi ^3}{r^3}\cos \theta }},\\
	CR{B_r} &\approx \frac{1}{{2\gamma L}}\frac{{{\lambda ^2}\left[{{\cos }^2}\theta {{(\frac{{{D_T}}}{{r\cos \theta }})}^2} + \frac{\pi{D_T}{\cos 2\theta  }}{r{\cos \theta}} - 4{{\sin }^2}\theta {{(\ln \frac{{{D_T}}}{{r\cos \theta }})}^2}\right]}}{{8{\pi ^2}M(\frac{{\pi {D_T}}}{r{\cos \theta}} - 4{{\ln }^2}\frac{{{D_T}}}{{r\cos \theta }})}}\\
	&\mathop  \approx \limits^{(g)} \frac{1}{{2\gamma }}\frac{{{\lambda ^2}{{(\frac{{{D_T}}}{r})}^2}}}{{8{\pi ^2}M\frac{{\pi {D_T}}}{r{\cos \theta}}}} \approx \frac{1}{{2\gamma }}\frac{{{\lambda ^2}{d_T}\cos \theta }}{{8{\pi ^3}r}},
	\end{aligned}
\end{equation}
where $(g)$ follows from$\ \mathop {\lim }\limits_{x \to \infty } \left( {a{x^2} + bx - c{{\ln }^2}x} \right) = a{x^2},\ $$\ \mathop {\lim }\limits_{x \to \infty } \left( {bx - c{{\ln }^2}x} \right) = bx,\ $with $x=\frac{D_T}{r\cos\theta}$.
%
%When $\theta=0$, we have 
%\begin{equation}
%	\setlength\abovedisplayskip{1.5pt}
%\setlength\belowdisplayskip{1.5pt}
%	\footnotesize
%	\begin{aligned}
%		%			\ \begin{array}{*{20}{l}}
%		a &\approx \frac{{4{\pi ^2}{r^2} }}{{{\lambda ^2}{\varepsilon _T}}} ({\frac{{{D_T}}}{r} -\pi} ),
%		e \approx 0,c\approx0,\\
%		p &\approx \frac{{4{\pi ^2}}}{{{\lambda ^2}{\varepsilon _T}}}\pi,
%		q \approx j\frac{{4\pi }}{{\lambda {\varepsilon _T}}}(\ln \frac{{{D_T}}}{{r }}).
%		%			\end{array}\
%	\end{aligned}
%\end{equation}
%Therefore, the CRBs can be expressed as
%\begin{equation}
%	\setlength\abovedisplayskip{1.5pt}
%	\setlength\belowdisplayskip{1.5pt}
%	\footnotesize
%	\begin{aligned}
%		CR{B_\theta } &\approx \frac{1}{2{T_p}{\gamma }}\frac{{{\lambda ^2}}}{{4M{\pi ^2}{r^2}}},\\
%		CR{B_r} &\approx \frac{1}{{2{T_p}{\gamma }}}\frac{{{\lambda ^2}( {\frac{{D_T^2}}{{{r^2}}} + \pi \frac{{{D_T}}}{r}} )}}{{4{\pi ^2}( {\pi \frac{{{D_T}}}{r} - 4{{\ln }^2}\frac{{{D_T}}}{r}} )}}
%		\approx \frac{1}{{2\gamma }}\frac{{{\lambda ^2}{d_T}}}{{8{\pi ^3}r}}.\
%	\end{aligned}
%\end{equation}
\section*{Appendix D\\Proof of Corollary \ref{jinsifar}}
Similar to the derivation in Appendix B, when $D_T/r\ll 1$, the functions $g_1,~g_2,~g_3,~g_4$ in Appendix B reduce to
\begin{equation}
	\setlength\abovedisplayskip{1.5pt}
	\setlength\belowdisplayskip{1.5pt}
\footnotesize
	\begin{aligned}
	{g_1}\left( {\frac{{{D_T}}}{r}} \right)&\mathop  \approx \limits^{(h1)}  - \frac{{{D_T}}}{r}\sin \theta,~
	{g_2}\left( {\frac{{{D_T}}}{r}} \right)\mathop  \approx \limits^{(h2)}  \frac{{{D_T}}}{r}\cos \theta - \frac{{{{\cos }^3}\theta }}{{12}}{\left( {\frac{{{D_T}}}{r}} \right)^3},\\
	{g_3}\left( {\frac{{{D_T}}}{r}} \right)&\mathop  \approx \limits^{(h3)}  - 2 \frac{{{D_T}}}{r}\sin \theta + \frac{{\sin \theta }}{2}{\left( {\frac{{{D_T}}}{r}} \right)^3},~
	{g_4}\left( {\frac{{{D_T}}}{r}} \right)\mathop  \approx \limits^{(h4)} \frac{{{D_T}}}{r},
	\label{g1-4}
	\end{aligned}
\end{equation}
where $(h1)$ and $(h4)$ follow from the first-order Taylor approximation, and $(h2)$ and $(h3)$ follow from the third-order and second-order Taylor approximation. Note that $(h1)$ and $(h4)$ are related to $c$ and $q$, which only appear in the form of square, so there is no need to consider higher order approximation when $D_T/r\ll 1$. By substituting (\ref{g1-4}) into (\ref{parametera})-(\ref{parameterq}), the closed form expressions of intermediate parameters $a,e,p,c,q$ are obtained. 
%Furthermore, some major terms in (\ref{crb-mimo1}) can be obtained as
%\begin{equation}
%	\setlength\abovedisplayskip{1.5pt}
%	\setlength\belowdisplayskip{1.5pt}
%	\footnotesize
%	\begin{aligned}
%	&Mp - |q{|^2} \approx  - \frac{{4{\pi ^2}M{{\cos }^2}\theta }}{{12{\lambda ^2}{\varepsilon _T}}}\left( {6{{\sin }^2}\theta  + {{\cos }^2}\theta \cos 2\theta } \right){\left( {\frac{{{D_T}}}{r}} \right)^3},\\
%	&(Ma - |c{|^2})(Mp - |q{|^2}) - {[Me - {c^*}q]^2} \approx  -\left( {9{{\sin }^2}\theta  + {{\cos }^6}\theta } \right)\\
%	& \times{\left( {\frac{{4{\pi ^2}rM\cos \theta }}{{12{\lambda ^2}{\varepsilon _T}}}} \right)^2}{\left( {\frac{{{D_T}}}{r}} \right)^6}.
%		\label{farint}
%	\end{aligned}
%\end{equation}
By substituting (\ref{g1-4}) into (\ref{crb-mimo1}), the CRB expression in (\ref{jinsifar1}) can be obtained.
\section*{Appendix E\\Proof of Corollary \ref{lem4}}
The second-order Taylor approximation of the distance between the target and the $m$th antenna element in (\ref{eq9}) is
\begin{equation}
	\setlength\abovedisplayskip{1.5pt}
\setlength\belowdisplayskip{1.5pt}
\footnotesize
\begin{aligned}
	{r_m} \approx r + \frac{1}{{2r}}{( {m{d_T}} )^2}{\cos ^2}\theta  - m{d_T}\sin \theta .\
	\label{taylor}
	\end{aligned}
\end{equation}
Therefore, the element of the steering vector ${a_m}(\theta ,r)$ in (\ref{eq2}) can be expressed as~${{a_m}(\theta ,r) = {e^{jmv + j{m^2}\phi  - j\frac{{2\pi }}{\lambda }r}}},\ $where$\ v = \frac{{2\pi {d _T}\sin \theta }}{\lambda }$,~$\phi  =  - \frac{{\pi {{d _T}^2}{{\cos }^2}\theta }}{{\lambda r}}.\ $The derivatives of angle and range of steering vector are
\begin{equation}
			\setlength\abovedisplayskip{0pt}
\setlength\belowdisplayskip{1pt}
\footnotesize
\begin{aligned}
			\frac{{\partial {a_m}(r,\theta )}}{{\partial \theta }} &= {e^{jmv + j{m^2}\phi  - j\frac{{2\pi }}{\lambda }r}}\left( {jm\frac{{\partial v}}{{\partial \theta }} + j{m^2}\frac{{\partial \phi }}{{\partial \theta }}} \right),\\
	\frac{{\partial {a_m}(r,\theta )}}{{\partial r}} &= {e^{jmv + j{m^2}\phi  - j\frac{{2\pi }}{\lambda }r}}\left( {j{m^2}\frac{{\partial \phi }}{{\partial r}} - j\frac{{2\pi }}{\lambda }} \right),
\end{aligned}
\end{equation}
where $\frac{{\partial v}}{{\partial \theta }} = \frac{{2\pi {d_T}\cos \theta }}{\lambda },\frac{{\partial \phi }}{{\partial \theta }} = \frac{{\pi {d_T}^2\sin 2\theta }}{{\lambda r}},\ $and$\ \frac{{\partial \phi }}{{\partial r}} = \frac{{\pi {d_T}^2{{\cos }^2}\theta }}{{\lambda {r^2}}}.$
Therefore, the intermediate parameters $a,c,e,p,q$ based on the second-order Taylor approximation model can be derived as
\begin{equation}
				\setlength\abovedisplayskip{1pt}
	\setlength\belowdisplayskip{1pt}
\scriptsize
	\begin{aligned}
			a&={{{\left\| {\frac{{\partial {\bf{a}}(r,\theta )}}{{\partial \theta }}} \right\|}^2} = \sum\nolimits_{m =  - \frac{{M - 1}}{2}}^{\frac{{M - 1}}{2}} {{m^2}{\left( {\frac{{2\pi {d _T}\cos \theta }}{\lambda } + m\frac{{\pi {{d _T}^2}\sin {2\theta }}}{{\lambda r}}}\right )^2}} }\\
			&= {\left( {\frac{{2\pi {d _T}\cos \theta }}{\lambda }} \right)^2}{\eta_1} + {\left( {\frac{{\pi {{d _T}^2}\sin {2\theta }}}{{\lambda r}}}\right )^2}{\eta_2},
	\end{aligned}
\end{equation}
where$\ {\eta_1} = \sum\nolimits_{m =  - \frac{{M - 1}}{2}}^{\frac{{M - 1}}{2}} {{m^2}}  = \frac{{M( {{M^2} - 1} )}}{12}\ $and$\ {\eta_2} = \sum\nolimits_{m =  - \frac{{M - 1}}{2}}^{\frac{{M - 1}}{2}} {{m^4}}  = \frac{{M( {{M^2} - 1} )( {3{M^2} - 7} )}}{{240}}.\ $Similarly, we can obtain other parameters:
\begin{equation}
\setlength\abovedisplayskip{2pt}
\setlength\belowdisplayskip{2pt}
\footnotesize
\begin{aligned}
	p&= {{\left( {\frac{{\pi {{d _T}^2}{{\cos }^2}\theta }}{{\lambda {r^2}}}} \right)}^2}{\eta_2} - {{\left( {\frac{{2\pi {d _T}\cos \theta }}{\lambda r }}\right )}^2}{\eta_1} + \frac{{4{\pi ^2}}}{{{\lambda ^2}}}M,\\
	e&= \left( {\frac{{{\pi ^2}{{d _T}^4}{{\cos }^2}\theta \sin 2\theta }}{{{\lambda ^2}{r^3}}}} \right){\eta_2} - \frac{{2{\pi ^2}{{d _T}^2}\sin 2\theta }}{{{\lambda ^2}r}}{\eta_1},\\
	c&=  - j\frac{{\pi {{d _T}^2}\sin2\theta }}{{\lambda r}}{\eta_1},
	q=  - j\frac{{\pi {{d _T}^2}{{\cos }^2}\theta }}{{\lambda {r^2}}}{\eta_1} + j\frac{{2\pi }}{\lambda }M.
		\label{para_tay}
	\end{aligned}
\end{equation}
%Furthermore, the major terms in (\ref{crb-mimo1}) and (\ref{crb-mimo2}) are
%\begin{equation}
%	\setlength\abovedisplayskip{2pt}
%	\setlength\belowdisplayskip{2pt}
%	\footnotesize
%	\begin{aligned}
%		&Me - {c^*}q = \left( {\frac{{{\pi ^2}{{d _T}^4}{{\cos }^2}\theta \sin 2\theta }}{{{\lambda ^2}{r^3}}}}\right )( {M{\eta_2} - {\eta_1}^2} ),\\
%		&Mp - {{| q |}^2} = {{\left( {\frac{{\pi {{d _T}^2}{{\cos }^2}\theta }}{{\lambda {r^2}}}} \right)}^2}( {M{\eta_2} - {\eta_1}^2} ),\\
%		&( {Ma - {{| c |}^2}} ) - \frac{{{{[ {Me - {c^*}q} ]}^2}}}{{( {Mp - {{| q |}^2}} )}} = M{{\left( {\frac{{2\pi {d _T}\cos \theta }}{\lambda }} \right)}^2}{\eta_1}.
%			\label{1}
%	\end{aligned}
%\end{equation}
By substituting (\ref{para_tay}) into (\ref{crb-mimo1})-(\ref{crb-mimo2}), the expressions in (\ref{crb-taylor-mimo1}) and (\ref{crb-taylor-mimo2}) can be obtained accordingly.

\section*{Appendix F\\Proof of Corollary \ref{bistatic}}
When $\theta=0$, the parameters in (\ref{gamma}) reduce to
\begin{equation}
	\setlength\abovedisplayskip{1.5pt}
	\setlength\belowdisplayskip{1.5pt}
	\footnotesize
	\begin{aligned}
		{\Gamma _\theta }(r,\theta ) = \frac{r}{{R - r}},~{\Gamma _r}(r,\theta ) = 0.
	\end{aligned}
\end{equation}
Therefore, the parameters in (\ref{bistatic1}) and (\ref{bistatic2}) reduce to
\begin{equation}
	\setlength\abovedisplayskip{1.5pt}
	\setlength\belowdisplayskip{1.5pt}
	\footnotesize
	\begin{aligned}
		a &= \frac{{4{\pi ^2}{r^2}}}{{{\lambda ^2}{\varepsilon _T}}}\left[ {\frac{{{D_T}}}{r}} \right.\left. { - 2\arctan \left( {\frac{{{D_T}}}{{2r}}} \right)} \right],~c = e = s = k = f = h = 0,\\
		p &= \frac{{4{\pi ^2}}}{{{\lambda ^2}{\varepsilon _T}}}2\arctan \left( {\frac{{{D_T}}}{{2r}}} \right),~q = j\frac{{2\pi }}{{\lambda {\varepsilon _T}}}\ln \left( {\frac{{\frac{{{D_T}}}{{2r}} + \sqrt {1 + {{\left( {\frac{{{D_T}}}{{2r}}} \right)}^2}} }}{{ - \frac{{{D_T}}}{{2r}} + \sqrt {1 + {{\left( {\frac{{{D_T}}}{{2r}}} \right)}^2}} }}} \right),\\
		i &= \frac{{{\pi ^2}d_R^2{r^2}}}{{3{\lambda ^2}{{(R - r)}^2}}}N({N^2} - 1),
	\end{aligned}
\label{Fsimpli}
\end{equation}
Therefore, the CRBs in (\ref{bistatic1}) and (\ref{bistatic2}) reduce to
\begin{equation}
		\setlength\abovedisplayskip{1.5pt}
	\setlength\belowdisplayskip{1.5pt}
\footnotesize
\begin{aligned}
	CR{B_\theta } = \frac{1}{{2\gamma L}}\frac{M}{{Mi + Na}},	~CR{B_r} = \frac{1}{{2\gamma LN}}\frac{M}{{p - \frac{1}{M}{{\left| q \right|}^2}}}.
	\label{F2simpli}
\end{aligned}
\end{equation}
By substituting (\ref{Fsimpli}) in (\ref{F2simpli}), the closed-form expressions in (\ref{0bistatic1}) and (\ref{0bistatic2}) can be obtained. In order to obtain the minimal point of range CRB, the derivative of $CRB_r$ with respect to $\frac{D_T}{2r}$ is derived. Let $x=\frac{D_T}{2r}$, and the derivative can be expressed as
\begin{equation}
\setlength\abovedisplayskip{1pt}
\setlength\belowdisplayskip{1pt}
\footnotesize
\begin{aligned}
	&\frac{{\partial CR{B_r}}}{{\partial x}} = \frac{{{\lambda ^2}}}{{8{\pi ^2}\gamma LN}}\\
	&\times\frac{{\arctan x - \frac{x}{{1 + {x^2}}} + \frac{2}{{\sqrt {1 + {x^2}} }}\ln \left( {x + \sqrt {1 + {x^2}} } \right) - \frac{2}{x}{{\ln }^2}\left( {x + \sqrt {1 + {x^2}} } \right)}}{{{{\left[ {\arctan x - \frac{1}{x}{{\ln }^2}\left( {x + \sqrt {1 + {x^2}} } \right)} \right]}^2}}}
\end{aligned}
\end{equation}
In order to obtain the minimal point, let$\ \frac{{\partial CR{B_r}}}{{\partial x}} = 0\ $and solution is $x\approx6$ by numerical simulation, i.e., $D_T \approx 12r$.

%\section*{Appendix G\\Proof of Corollary \ref{th8}}
%When $\frac{D_T}{r\cos\theta}\gg1$, by substituting the $g_2$ and $g_4$ in Appendix B into (\ref{0bistatic1}) and (\ref{0bistatic2}), the closed-form expression in (\ref{inf-bistatic1}) and (\ref{inf-bistatic2}) can be obtained accordingly. Note that the CRB of range can be written as
%\begin{equation}
%	\setlength\abovedisplayskip{1.5pt}
%	\setlength\belowdisplayskip{1.5pt}
%\footnotesize
%	\begin{aligned}
%		CR{B_r}\mathop  \approx \limits^{(a)} \frac{1}{{2\gamma L}}\frac{{{D_T}^2{\lambda ^2}}}{{4{\pi ^2}{r^2}N\pi \frac{{{D_T}}}{r}}}	= \frac{1}{{2\gamma L}}\frac{{{\lambda ^2}}}{{4{\pi ^2}N\pi }}\frac{{{D_T}}}{r}\to\infty.
%	\end{aligned}
%\end{equation}
%where $(a)$ follows from Appendix C.
%\begin{equation}
%	
%\end{equation}
\bibliographystyle{IEEEtran}
\bibliography{IEEEabrv,ref}
\end{document}